\documentstyle[aaspp4,11pt,epsf,tighten]{article}
\begin{document}

\title{Globular Cluster Photometry with the Hubble Space Telescope.\\
VII. Color Gradients and Blue Stragglers in the Central Region\\
of M30 (NGC~7099) from WFPC2 Observations\footnote{Based on observations with
the NASA/ESA Hubble Space Telescope, obtained at the Space Telescope Science
Institute, which is operated by the Association of Universities for Research
in Astronomy, Inc., under NASA contract NAS5-26555.}}
\author{Puragra Guhathakurta\footnote{Alfred P.\ Sloan Research Fellow} and
Zodiac T.\ Webster}
\affil{UCO/Lick Observatory, Department of Astronomy \& Astrophysics,\\
University of California, Santa Cruz, California 95064, USA\\
Email: {\tt raja@ucolick.org}, {\tt zodiac@ucolick.org}}
\author{Brian Yanny}
\affil{Fermi National Accelerator Laboratory, Batavia, Illinois 60510, USA\\
Email: {\tt yanny@sdss.fnal.gov}}
\author{Donald P.\ Schneider}
\affil{Department of Astronomy \& Astrophysics, The Pennsylvania State
University,\\
University Park, Pennsylvania 16802, USA\\
Email: {\tt dps@astro.psu.edu}}
\and
\author{John N.\ Bahcall}
\affil{Institute for Advanced Study, Princeton, New Jersey 08540, USA\\
Email: {\tt jnb@sns.ias.edu}}

\begin{abstract}
We present F555W ($V$), F439W ($B$), and F336W ($U$) photometry of 9507~stars
in the central $2'$ of the dense, post core collapse cluster M30 (NGC~7099)
derived from {\it Hubble Space Telescope\/} Wide Field/Planetary Camera~2
images.  These data are used to study the mix of stellar populations in the
central region of the cluster.  Forty eight blue straggler stars are
identified; they are found to be strongly concentrated towards the cluster
center.  The specific frequency of blue stragglers, $F_{\rm
BSS}\equiv{N}({\rm BSS})/N(V<V_{\rm HB}+2)$, is~$0.25\pm0.05$ in the inner
region of M30 ($r<20''$), significantly higher than the frequency found in
other clusters: $F_{\rm BSS}=0.05\>$--$\>$0.15.  The shape of M30's blue
straggler luminosity function resembles the prediction of the collisional
formation model and is inconsistent with the binary merger model, of Bailyn
\& Pinsonneault (1995, ApJ, 439, 705).  An unusually blue star ($B=18.6$,
$B-V=-0.97$), possibly a cataclysmic variable based on its color, is found
about $1\farcs2$ from the crowded cluster center.  Bright red giant stars
($B<16.6$) appear to be depleted by a factor of~2--3 in the inner $r<10''$
relative to fainter giants, subgiants, and main sequence turnoff stars (95\%
significance).  We confirm that there is a radial gradient in the color of
the overall cluster light, going from $B-V\sim0.82$ at $r\sim1'$ to
$B-V\sim0.45$ in the central $10''$.  The central depletion of the bright red
giants is responsible for about half of the observed color gradient; the rest
of the gradient is caused by the relative underabundance of faint red main
sequence stars near the cluster center (presumably a result of mass
segregation).  The luminosity function of M30's evolved stars does not match
the luminosity function shape derived from standard stellar evolutionary
models: the ratio of the number of bright giants to the number of turnoff
stars in the cluster is 30\% higher than predicted by the model ($3.8\sigma$
effect), roughly independent of red giant brightness over the range $M_V=-2$
to $+2$.
\end{abstract}


\section{Introduction}

The {\it Hubble Space Telescope\/} ({\it HST\/}) is ideally suited for the
study of individual stars in the crowded central regions of dense Galactic
globular clusters.  The projected density of evolved stars alone ranges from
about 0.3~arcsec$^{-2}$ in the core of the well studied dense cluster M13 to
$\gtrsim30$~arcsec$^{-2}$ at the center of M15, the Galactic cluster with the
highest known central surface density (\cite{randi}, hereafter Paper~VI).
Resolving individual stars at these densities is beyond the current angular
resolution limits of ground-based telescopes, but such observations are
crucial in order to address several important astrophysical issues.

Globular clusters are excellent testing grounds for models of stellar
evolution because the cluster members are coeval and have similar chemical
composition (cf.~\cite{bergbvan}).  Deviations from the natural process of
(isolated) stellar evolution caused by stellar interactions in these
environments of extreme star density can be explored by examining the radial
dependence of the mix of stellar populations (cf.~\cite{piotto88};
\cite{djor91}).  Additionally, the dense cores of globular clusters are
unique laboratories for studying the effects of two-body relaxation,
equipartition of energy, and binaries on the dynamical evolution of dense
stellar systems (cf.~\cite{hut92}; \cite{meyl97}).  A visible product of
stellar collisions and mergers of binaries, blue straggler stars (BSSs), are
preferentially found in the central regions of most globular clusters
(\cite{ferrarom3}; \cite{bsstheory}).

This is the seventh in a series of papers describing {\it HST\/} observations
of the centers of the nearest Galactic globular clusters with
$\vert{b}\vert>15^{\circ}$.  The main scientific goals of this program are to
measure the shape of the density profile in clusters and to understand the
nature of evolved stellar populations in very dense regions by probing the
variation in the mix of stellar types as a function of radius (and hence
stellar density).  Complementary programs targeting main sequence (MS) stars
in globular clusters are being conducted independently by other groups to
explore cluster dynamics (cf.~\cite{massseg}) and models of stellar evolution
(cf.~\cite{PCK}).  In this paper we examine the evolved stellar populations
of M30 (NGC~7099) using a set of techniques, developed in earlier papers
(\cite{ppr1}; \cite{ppr2}; \cite{ppr5}, hereafter referred to as Papers~I,
II, and V, respectively), to:
(1)~Build empirical point spread function (PSF) models using isolated bright
  stars, allowing for PSF variability across the image and using faint stars
  to reconstruct the saturated cores of brighter stars;
(2)~Iteratively fit the PSF template to the stars on the image;
(3)~Perform aperture photometry on each star after subtracting its
  neighboring stars with the best-fit PSF template; and
(4)~Carry out detailed and realistic image simulations to assess the effects
  of crowding on photometric accuracy and sample completeness.

M30 is a prototypical ``post-core-collapse'' globular cluster based on its
rising surface brightness profile at radii~$<3\arcsec$.  Its central surface
brightness in the $V$ band is $\mu_V(0)=15.20$~mag~arcsec$^{-2}$
(\cite{metll}), corresponding to a projected density of $\sim5$~arcsec$^{-2}$
in post MS stars alone.  The cluster has a relatively low metallicity,
$\rm[Fe/H]=-2.13$ (\cite{djor93}; \cite{zinn}), a low line-of-sight
reddening, $E_{B-V}=0.05$ (\cite{burs}), and is located at a distance of
about 9.8~kpc (\cite{reid}).  Based on this recent distance determination,
the age of M30 is estimated to be about 10~Gyr (\cite{Sandquist}).  It has
long been noted that M30 displays a central bluing trend like other
post-core-collapse clusters (\cite{willbach}; \cite{chunfree}; \cite{cord};
\cite{peterson86}; \cite{piotto88}; \cite{burgbuat}).

In this paper we use {\it HST\/} Wide Field/Planetary Camera~2~(WFPC2) data
to investigate the origin of M30's nuclear color gradient and to study its
abundant population of BSSs; Yanny et~al.\ (1994b, hereafter \cite{letter})
used the same data set to examine the inner 20\arcsec\ of M30.  This paper is
organized as follows: Sec.~2 contains a description of the data and reduction
techniques; Sec.~3 contains results from our study of M30's evolved stellar
populations, with particular emphasis on BSSs, radial population gradients,
and the stellar luminosity function (LF); Sec.~4 contains a summary of the
main points of the paper.

\section{Data}
\subsection{Observations}

A set of eight~WFPC2 images of the center of M30 (NGC 7099) were obtained on
1994 March~31 (see Paper~IV for a discussion of the data set): $2\times100$~s
with the F336W filter, $2\times40$~s with F439W, and $4\times4$~s with F555W.
These filters are roughly similar to the Johnson $U$, $B$, and $V$
bandpasses, respectively.  All eight~exposures were obtained at the same
telescope pointing and roll angle.

\begin{figure} 
\plotone{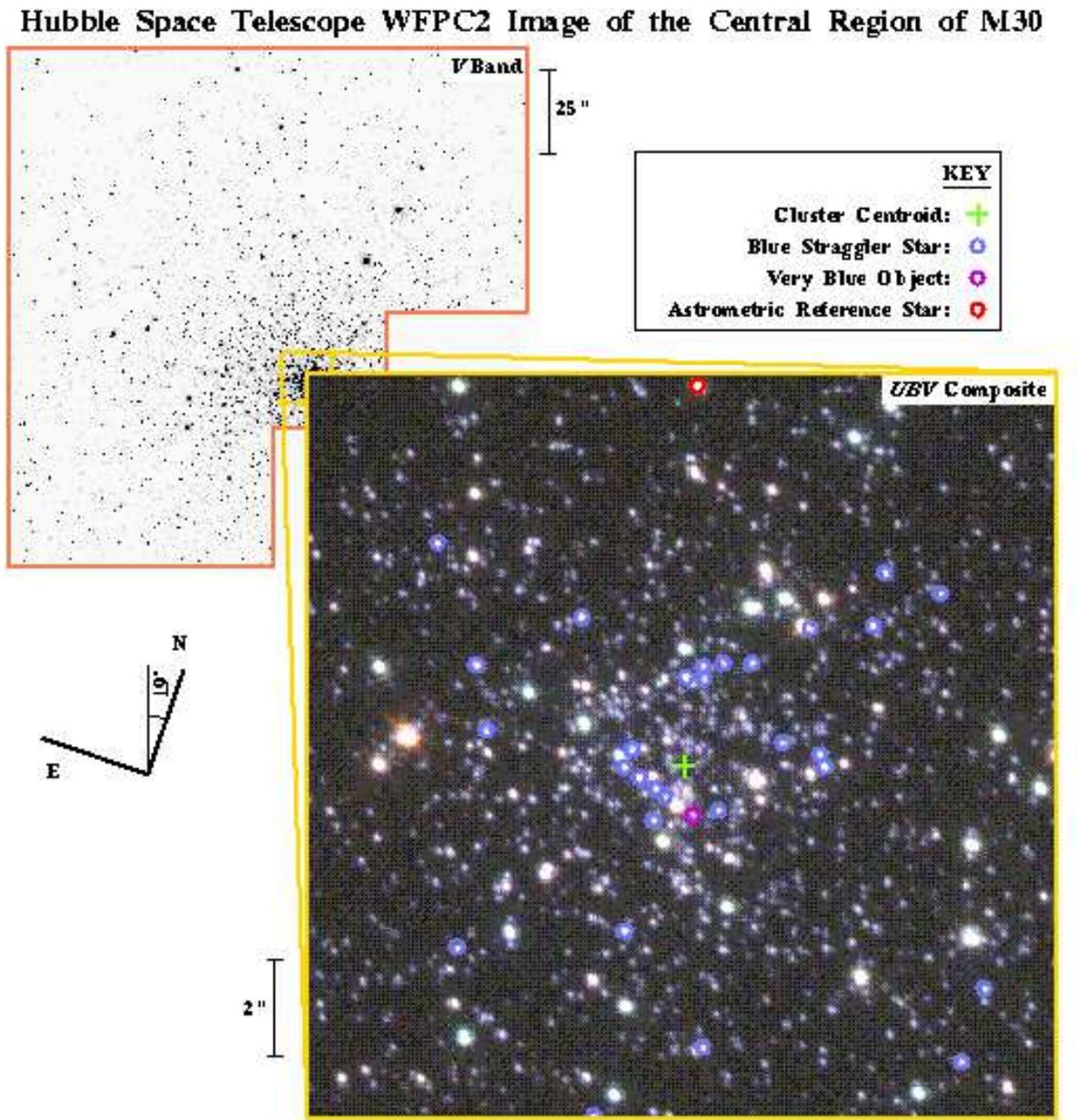}
\caption{A $V$-band WFPC2 image of M30 (negative
greyscale) and a ``true color'' $U+B+V$ composite of the cluster's central
$15''\times15''$ (color inset), along with the image scales and orientation.
The cluster centroid is indicated by the green `+'.  The colored circles
indicate: 29~blue stragglers (blue); a very blue star near the cluster center
(mauve); and the astrometric reference star near the top of the color inset
(red).  \label{pictm30}}
\end{figure}

Each WFPC2 image consists of four $800\times800$ Charge Coupled
Device~(CCD) frames: a Planetary Camera CCD~(PC1) with a scale of
$0\farcs0455$~pixel$^{-1}$ and three~Wide Field Camera (WF2--WF4) CCDs with a
scale of $0\farcs0996$~pixel$^{-1}$.
The usable field of view is $34\arcsec\times34\arcsec$ for PC1 and about
$76\arcsec\times76\arcsec$ for each WF CCD; the total image area is about
5.1~arcmin$^2$.  An image of the cluster is shown in Fig.~\ref{pictm30}
(Plate~XXX).  The telescope pointing was chosen so that the center of the
cluster was positioned near the center of the PC1 CCD, shown in the lower
right quadrant of the greyscale WFPC2 mosaic image.  The WF2--WF4 images are
oriented counterclockwise starting from the upper right quadrant of the
mosaic.  A detailed description of the instrumental parameters and in-orbit
characteristics of WFPC2 is given by \cite{burrows95} and by \cite{trau94}.

The exposure times have been chosen so that the brightest stars
($V\approx11.8$, $B\approx13.3$, $U\approx14.5$) in the image are not
saturated by more than a factor of~4--5 in each individual exposure.  In
general, no more than 10~pixels are affected per star, and it is possible to
recover their magnitudes and positions using information in the unsaturated
wings of the PSF.

\subsection{Data Processing / Stellar Photometry Technique}\label{dataproc}

We have developed and tested procedures for deriving stellar photometry from
{\it HST\/} images of dense globular clusters.  The reader is referred to
Papers~I and II for a detailed description.  Even though the procedures were
originally designed for pre-repair {\it HST\/} data with an aberrated PSF,
several of the steps are relevant for analysis of crowded star fields and we
apply them here.  The following steps for processing WFPC2 data are
summarized below (see \cite{ppr5}): cosmic ray removal, image coaddition,
star finding, PSF building, PSF fitting/subtraction, and aperture photometry.

Standard bias and flat field calibrations have been applied to the data as
part of the preprocessing pipeline at the Space Telescope Science Institute.
Pixels affected by cosmic rays are identified and masked using an algorithm
that compares each image to the other image(s) in that  band; unmasked
pixels in all images in a given band are averaged to obtain the final F336W,
F439W, and F555W images.

Cross-correlating pairs of images in different bands shows that the stellar
positions in the F555W and F439W band PC1 images are offset from the
corresponding stellar positions in the F336W band PC1 image by 1.2~pixel
($0\farcs055$) and 0.7~pixel ($0\farcs032$), respectively.  For the WF2--WF4
CCDs, the offset is in the range of 0.34$\>$--$\>$0.59~pixel
($0\farcs034$--$0\farcs059$) between the F555W and F336W images and in the
range 0.15$\>$--$\>$0.32~pixel ($0\farcs015$--$0\farcs032$) between the F439W
and F336W images.  The combined F336W, F439W, and F555W images are aligned by
fractional pixel interpolation and summed to produce a deep image.
Interpolation results in degradation of the PSF by an amount that varies from
star to star due to the undersampling of the PC and particularly the WF CCD
data; however, the interpolated image is only used to find stars (see below)
and {\it not\/} to derive stellar photometry.

A preliminary list of stars is obtained by applying the standard peak finding
algorithm {\sc find} of the {\sc daophot} package (\cite{daophot}) to the
summed F336W+F439W+F555W image.  Bright relatively isolated stars are used to
build an empirical, quadratically variable {\sc daophot~ii} PSF template
(\cite{daophotii}) for each of the four~bands and for each of the four~CCD
frames (PC1, WF2--WF4).  This PSF template is fit to and subtracted from the
stars on the image using {\sc daophot~ii's allstar} routine.  After
inspecting the residual images (original minus best-fit template), the star
list is edited to remove spurious objects, such as PSF artifacts around
bright stars and hot pixels (the slight offset between stellar positions in
F336W and F555W images facilitates the identification of fixed pattern
artifacts), and to add stars fainter than the {\sc find} threshold.  The
process of PSF building, fitting, and subtraction is iterated a few times; in
each instance, the results of the previous iteration are used to remove
neighbors of the bright stars used for PSF building.  Finally, aperture
photometry is obtained for each star, after its neighbors have been removed
using the best-fit PSF template.  The photometry lists in the
three~bandpasses are merged by positional matching.  The final list of stars
identified and matched in each of the three~filters consists of 3114~stars
from the PC1 image (0.32~arcmin$^2$), and 2055, 1495, and 2843~stars from
WF2--WF4, respectively (each WF CCD covers 1.60~arcmin$^2$).

Tests based on bright, relatively isolated stars show that the PSF is stable
across the WFPC2 field of view (see also Papers~IV and V): the correction
term for converting $r\sim0\farcs1$ aperture magnitudes to $r=1''$ ``total''
magnitudes varies by a few percent or less across each CCD image.
Furthermore, this slight variation is similar in all three~bands and the
telescope pointing is identical for all exposures in all bands so the effect
on measurement of stellar colors is negligible.

Differences in the zeropoints of the magnitude scale of the individual CCDs
are determined by cross-correlating the ``positions'' of bright
($V\lesssim18$) stars in the color-magnitude diagram~(CMD) between pairs of
CCD frames.  The tightness of the CMD features (Fig.~\ref{bvbcmd}), and the
fact that several hundred stars are used to define the peak of the
cross-correlation function, makes it possible to determine color and
magnitude offsets to an accuracy of~$\lesssim0.01$~mag. The inter-CCD
magnitude zeropoint offsets for WF2--WF4 with respect to P1 are [$+0.25$,
$+0.25$, $+0.25$] in F555W, [$+0.23$, $+0.22$, $+0.20$] in F439W, and
[$+0.43$, $+0.39$, $+0.41$] in F336W.  The zeropoint difference between the
WF CCDs and PC1 results primarily from the different aperture sizes used to
derive stellar photometry: $r_{\rm WF}=0\farcs16$ vs $r_{\rm PC1}=0\farcs09$.
The zeropoint difference is largest in F336W because its PSF core is
significantly broader than in F439W and F555W.

The cross-correlation technique used for inter-CCD zeropoint matching, as
well as the procedure used to match the WFPC2 photometry to ground-based
photometry (see below), relies on the validity of an implicit assumption:
That there are no intrinsic spatial gradients in the color of the individual
evolved stellar populations in M30, particularly in the $B-V$ and $U-V$
colors of faint red giant branch~(RGB) stars which dominate by number the
$V\lesssim18$ sample used for matching.  Radial gradients in RGB color have
never been convincingly demonstrated to exist in globular clusters; it is
common to find changes in the {\it mix\/} of stellar types, but not in the
color of a {\it given\/} stellar type.  Moreover, it is reassuring that there
is a perfectly plausible explanation for the observed inter-CCD zeropoint
differences in the M30 data set.

A charge transfer efficiency problem during WFPC2 readout (Burrows 1995;
\cite{holtz95b}) causes objects at high row number to appear systematically
fainter than if they had been at low row number.  For exposures taken at
$-77^\circ$C, a linear ramp correction ranging from no correction at row~1 to
$-0.1$~mag at row~800 (original CCD coordinates) works well to first order.
We apply such a linear ramp correction to the stellar photometry; it results
in a noticeable tightening of the horizontal branch~(HB) and subgiant
branch/main sequence turnoff~(MSTO), the only features that are more or less
horizontal in the CMD.  Note that color measurements are unaffected by the
charge transfer efficiency problem since the telescope pointing was identical
for exposures in all three~bands.   

The zeropoints of the WFPC2 instrumental F336W, F439W, and F555W magnitudes
are adjusted to roughly match ($\approx0.1$~mag) ground-based photometry of
stars in M30 in the Johnson $U$, $B$, and $V$ bands (\cite{mags}).  There is
not sufficient overlap between this ground-based $UBV$ data set (or any other
published ones) and the {\it HST\/} data set to allow an extensive
star-by-star comparison, the cluster's central region being too crowded at
arcsecond resolution.  Instead, various fiducial points in the CMD---HB
brightness, and the RGB color at the level of the HB and at the RGB
base---and a variety of filter combinations [(\bv,~$V$) and (\ub,~$B$)] are
used to determine the zeropoints in all three~bands.  These zeropoints are in
good agreement with the published WFPC2 photometric zeropoints (Burrows
1995).  Following the prescription of \cite{holtz95a}, the zeropoint-adjusted
magnitudes on the WFPC2 bandpass system, $m_{336}$, $m_{439}$, and $m_{555}$,
are converted to the Johnson $UBV$ system:

\[
U~=~m_{336}\,-\,0.240\,(m_{336}-m_{555})\,+\,0.048\,(m_{336}-m_{555})^2
\]
\[
B~=~m_{439}\,+\,0.003\,(m_{439}-m_{555})\,-\,0.088\,(m_{439}-m_{555})^2
\]
\begin{equation}
V~=~m_{555}\,-\,0.060\,(m_{439}-m_{555})\,+\,0.033\,(m_{439}-m_{555})^2
\end{equation}

\noindent
The $B$ and $V$ magnitude conversions are accurate, but the converted ``$U$''
magnitude may be significantly different from true Johnson $U$ because of the
severe red leak in the F336W bandpass (e.g.,~see Fig.~\ref{ubucmd}) which is
not adequately accounted for by quadratic terms in the color transformation
equation.

\subsection{Astrometry}\label{astrom}

Astrometric calibration has been carried out using the {\sc metric} task
(\cite{metric}) within {\sc iraf}'s {\sc stsdas} package.  This task uses
{\it HST\/} pointing information based on the Guide Star Catalog to derive
absolute astrometry, and corrects for mechanical geometric distortion and
spherical aberration that vary from CCD to CCD, to place all the stars on a
global coordinate system.  A relatively isolated bright star, ID\#~3611 in
Table~\ref{datatable} ($V=15.51$, $B=16.14$, $U=16.00$), is adopted as the
astrometric reference point:

\begin{equation}
\alpha_{\rm REF}({\rm J2000})=21^{\rm h}40^{\rm m}22\fs402,~~~
\delta_{\rm REF}({\rm J2000})=-23^{\circ}10'42\farcs42~~~.
\end{equation}

\noindent
The astrometric reference star (red circle in Fig.~\ref{pictm30}) is located
$2\farcs3$~E and $7\farcs6$~N of the cluster centroid (green cross in
Fig.~\ref{pictm30}).

The number-weighted centroid of evolved stars in M30, as determined in
\cite{letter}, has coordinates of: 

\begin{equation}
\alpha_0({\rm J2000})=21^{\rm h}40^{\rm m}22\fs24,~~~
\delta_0({\rm J2000})=-23^\circ10'50\farcs0~~~.
\end{equation}

\noindent
to a $1\sigma$ accuracy of $\delta{r}=0\farcs2$.  We adopt this as the
cluster center in the rest of the analysis presented in this paper.

\subsection{Photometric Accuracy and Completeness}\label{acccomp}

Using artificial star tests and realistic simulations of {\it HST\/} WFPC2
images of M30 and M15 carried out in Papers~IV and~V, respectively, we
estimate the degree of completeness and level of {\it internal\/} photometric
accuracy in the data set analyzed in this paper.  These error estimates, as
well as the other error estimates discussed below, refer to the internal
accuracy in the measurement of relative instrumental magnitudes; there may be
additional systematic errors in the conversion to Johnson magnitudes
($\lesssim0.1$~mag in $B$ and $V$, and significantly larger in $U$ due to the
F336W red leak).  The ratio of exposure times for M30 and M15 has been chosen
to compensate for the 0.76~mag difference in their distance moduli
(\cite{metll}) so that the ratio of signal to (read+Poisson) noise is the
{\it same\/} for stars of a given absolute magnitude.  The surface density of
post MS stars is about 10~times higher in M15 than in M30 (at comparable
angular distances from the cluster center), while the shape of the LF of
evolved stars is similar in the two clusters.  Thus, the effect of crowding
is slightly less severe in the M30 data set than in the M15 data
set---i.e.,~the degree of completeness is higher and the photometry more
accurate in M30 at a given absolute magnitude and angular distance from the
cluster center.

Our best estimate for the $1\sigma$~photometric error in
$V$ in the dense central $10''$ of M30 is~$\lesssim0.05$~mag for bright stars
($V\lesssim17$) and about 0.10$\>$--$\>$0.15~mag for stars at the brightness
level of the MSTO ($V\sim18.6$).  In this region of M30, effects of blending
are expected to start becoming noticeable around the MSTO.  The sample should
be complete down to the MSTO even in the crowded central regions of the
cluster.  The simulations also indicate that, within $r<5\arcsec$, the level
of completeness starts falling off for stars just below the MSTO
($V\gtrsim19$), dropping to roughly 50\% for stars 2~mag below the MSTO.
Incompleteness is expected to set in at progressively fainter magnitudes for
samples at larger radii.

We carry out a consistency check of these completeness estimates by studying
the observed LF of M30 stars as a function of radius.  The stars detected in
the full WFPC2 mosaic are divided into eight~radial bins around the cluster
center:
(1)~$r<5\farcs00$, 
(2)~$5\farcs00\leq{r}<9\farcs80$, 
(3)~$9\farcs80\leq{r}<15\farcs41$, 
(4)~$15\farcs41\leq{r}<23\farcs2$, 
(5)~$23\farcs2\leq{r}<35\farcs8$, 
(6)~$35\farcs8\leq{r}<51\arcsec$, 
(7)~$51\arcsec\leq{r}<71\arcsec$, and 
(8)~$71\arcsec\leq{r}<130\arcsec$. 
The boundaries of these radial bins are chosen so that each bin contains
roughly the same number of stars with $m_{555}<18.6$ ($\approx{V}_{\rm
MSTO}$); the data set should be complete at all radii for this bright
subsample of stars so it provides a convenient means of matching the relative
normalization of the LFs in the various radial bins.  The shape of the LF of
post MS stars is expected to be the same at all radii.  This is because stars
brighter than $M_V=M_V^{\rm MSTO}=+3.5$ lie within a very narrow mass range
($\Delta{M}\approx0.03\,M_\odot$---\cite{bergbvan}) so that the effects of
mass segregation are negligible (cf.~\cite{pryor86}; \cite{bolte89}).

\begin{figure}
\plotone{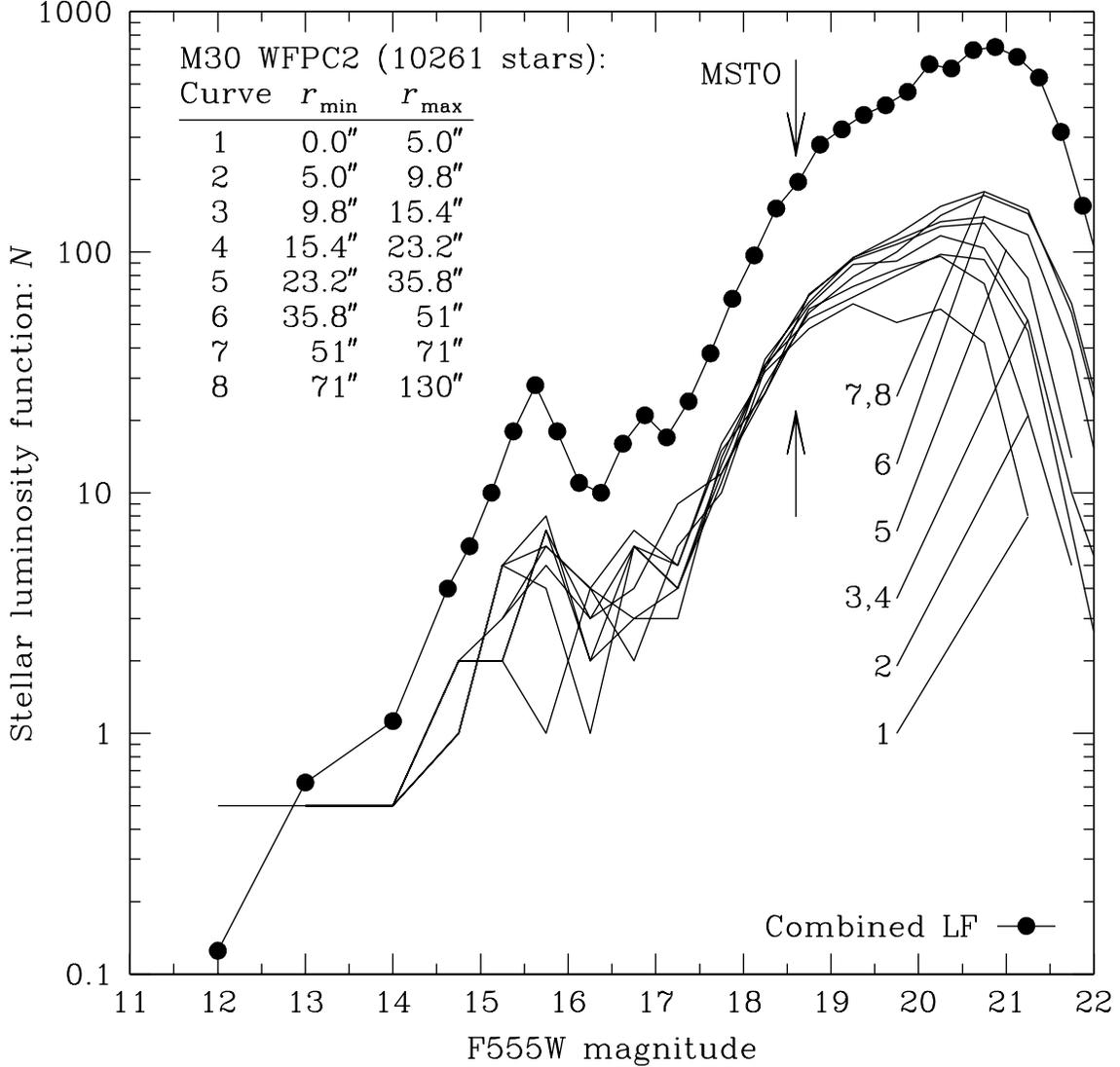}
\caption{The observed stellar luminosity
function of M30 in the F555W band (roughly Johnson $V$) in eight~radial bins.
The rollover beyond $\rm F555W\sim21$ is caused by incompleteness, which sets
in at brighter magnitudes at smaller radii because of increased crowding.
The bold solid curve with dots represents the ``most complete'' combined LF
based on the average all radial bins for $\rm F555W<18$ (to minimize Poisson
error) and on the average of the outermost two bins for stars with $\rm
F555W>18$ (since the degree of completeness is highest in this sparse outer
region).  The main sequence turnoff (MSTO) at $\rm F555W=18.6$ is marked; the
bump at $\rm F555W\sim15\>$--$\>$16 is caused by stars in the HB.
\label{vlum}}
\end{figure}

\begin{figure}
\plotone{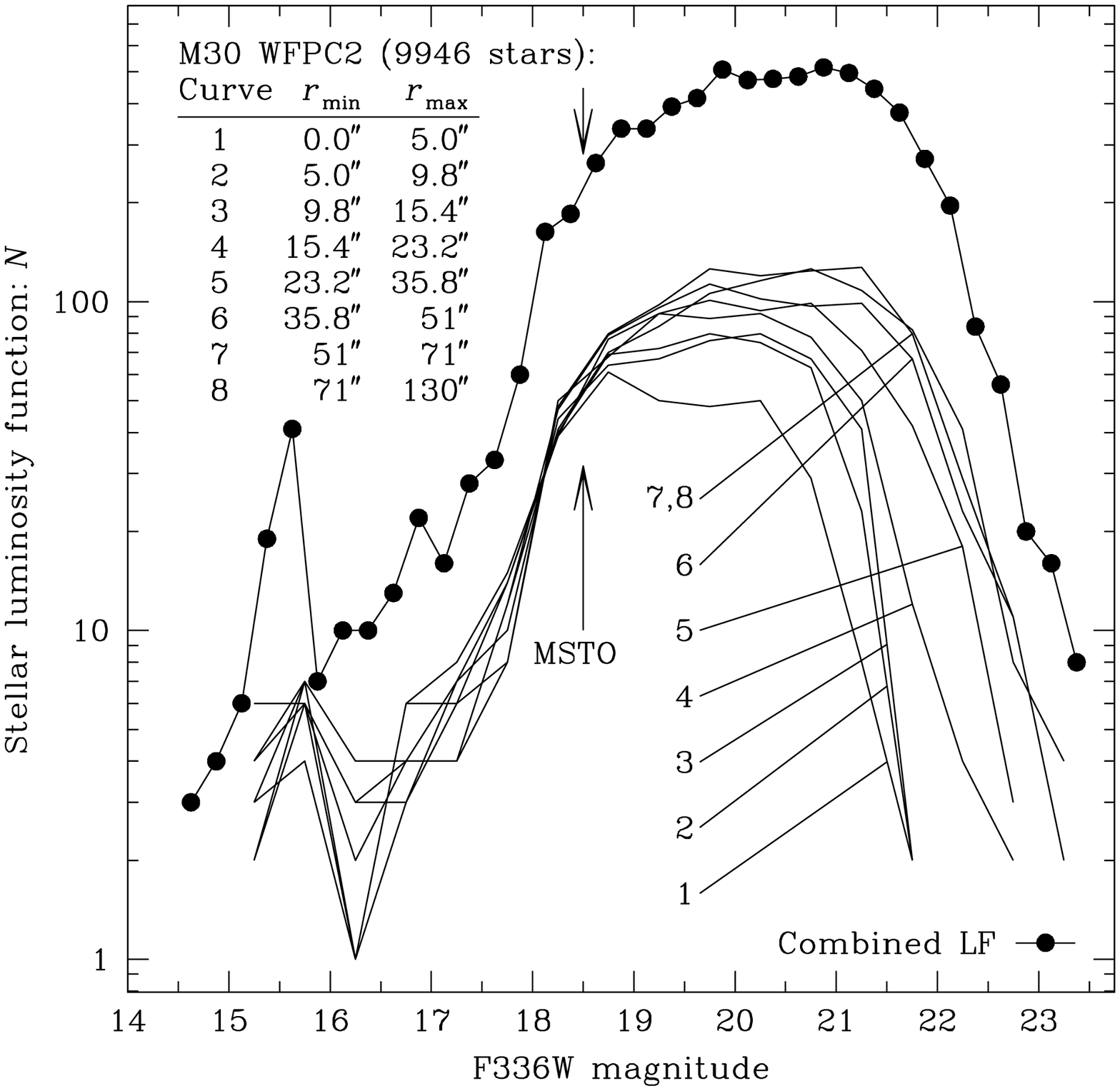}
\caption{Same as Fig.~\ref{vlum} for the F336W band
(roughly Johnson $U$).  The rollover beyond $\rm F336W\sim22$ is caused by
incompleteness.  As in Fig.~\ref{vlum}, the ``most complete'' combined LF
(bold solid line with dots) is based on the average of all bins for $\rm
F336W<18$ and on the average of the outermost two bins for $\rm F336W>18$.
The main sequence turnoff (MSTO) at $\rm F336W=18.5$ is marked; the bump at
$\rm F336W=15.5$ is produced by HB stars.  The LF slope in the RGB region is
steeper in the $U$ band than in $V$, while the MS slope is shallower, as
expected from the orientation of the RGB and MS tracks in the CMD.
\label{ulum}}
\end{figure}

The F555W-band stellar LFs are shown in Fig.~\ref{vlum}.  Note, these LFs are
based on stars detected on the F555W image, without requiring that the stars
also be detected in F439W or F336W; hence Johnson $V$ magnitudes are not
available for all these stars.  The bold line with dots shows the ``most
complete'' combined LF based on the average of all radial bins for $V<18$ (to
minimize Poisson error) and on the average of the two outermost annuli
($r\geq51\arcsec$) for stars with $V>18$ (since the degree of completeness is
highest in this sparse outer region).  The LF in the innermost bin
($r\lesssim5\arcsec$) is identical to the combined LF down to the MSTO, but
peels away at fainter magnitudes as expected, dropping sharply beyond
$V\sim20.5$ as a result of incompleteness.  The $V$ magnitude at which
incompleteness sets in appears to increase roughly monotonically with
increasing annular radius in keeping with the trend seen in the simulations
described above.

The F336W-band stellar LFs shown in Fig.~\ref{ulum} have a qualitatively
similar behavior to those in the F555W band.  A notable difference between
the two bands is that RGB stars span a narrower range of brightness in $U$
(3.5~mag) than in $V$ (5$\>$--$\>$6~mag), a natural consequence of the fact
that bright RGB stars tend to be redder than faint ones.  Thus, the
brightness contrast between faint MSTO stars and the brightest stars (at the
tip of the RGB) is smaller at the shorter wavelength, so MSTO stars are
slightly less affected by crowding in $U$ than in $V$.  This effect is partly
offset by the fact that the signal-to-noise ratio for faint MS stars is
slightly higher in $V$ than in $U$: the longer exposure time is not enough to
compensate for the lower instrumental efficiency and fainter intrinsic
stellar brightness in $U$.  Another difference between the LFs in the
two~bands is that the slope of the MS portion of the LF is shallower in $U$
than in $V$; this is because faint MS stars tend to be redder than bright
ones.  We return to a detailed discussion of the stellar LF in
Sec.~{\ref{lf_vs_model}}.

The widths of prominent features in different CMDs (Figs.~\ref{bvbcmd},
\ref{uvbcmd}, and \ref{ubucmd})---the RGB at $B=15\>$--$\>17$ and the MS just
below the MSTO ($B=19\>$--$\>20$)---are used to estimate the overall error in
stellar photometry, under the assumption that the features have zero
intrinsic width.  These error estimates represent averages over the entire
area of the WFPC2 mosaic.  For the filter combinations, $U-B$, $B-V$, and
$U-V$, the rms scatter in RGB color about empirically defined fiducial lines
is 0.04, 0.02, and 0.04~mag, respectively, and the rms scatter in MS color is
0.08, 0.08, and 0.08~mag, respectively.  These translate to photometric
errors (in magnitudes) of $\sigma_U=0.03$, $\sigma_B=0.01$, and
$\sigma_V=0.02$ for the bright stars comprising the RGB, and $\sigma_U=0.05$,
$\sigma_B=0.06$, and $\sigma_V=0.05$ for faint MS stars.

\section{Stellar Populations}

Our analysis of M30's stellar populations is based on photometry derived from
the {\it HST\/} WFPC2 images of the central region of the cluster.
Table~\ref{datatable} lists the relative positions, $\rm\Delta\alpha(J2000)$
and $\rm\Delta\delta(J2000)$, projected radial distance from the cluster
centroid, and Johnson $UBV$ magnitudes and colors of the 40~brightest stars
located within $r<10''$ of M30's center.  Stellar positions are on the
equinox J2000 coordinate system and are measured in arcseconds relative to
the astrometric reference star, ID\#~3611, whose coordinates are given in
Eq.~(2) in Sec.~\ref{astrom}.  The coordinates of the (number-weighted)
cluster centroid are given in Eq.~(3).  The complete electronic version of
Table~\ref{datatable} is available in the electronic edition of the Journal
by link to a permanent database and via {\tt anonymous ftp} (see Appendix);
it contains 9940~stars from the full M30 WFPC2 data set.  The stellar ID\#s
run sequentially through the full table in order of increasing right
ascension.

\begin{figure}
\plotone{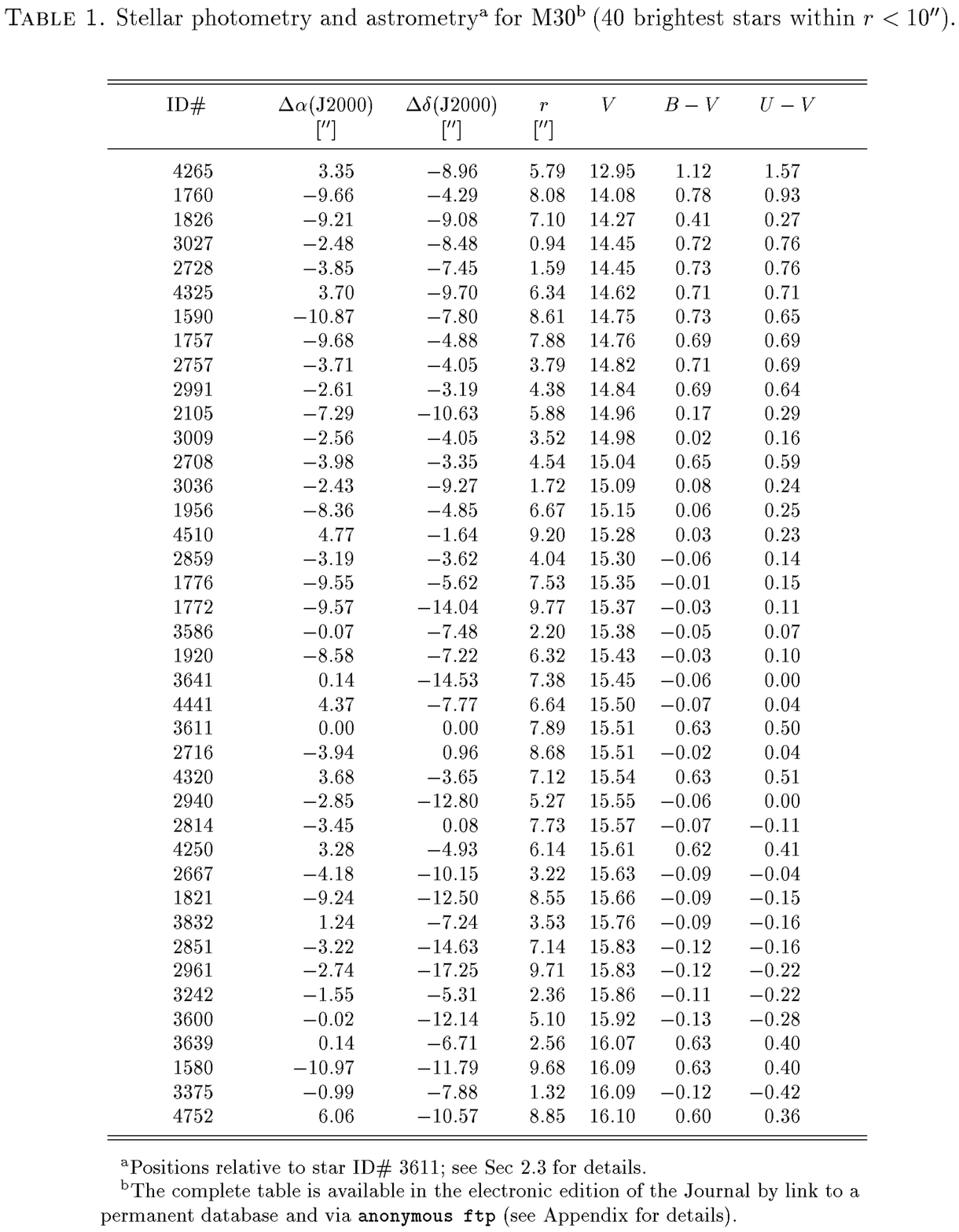}
\end{figure}
\placetable{datatable}
\begin{table}\dummytable\label{datatable}\end{table}

\subsection{Color-Magnitude Diagrams}\label{cmds}

\begin{figure}
\plotone{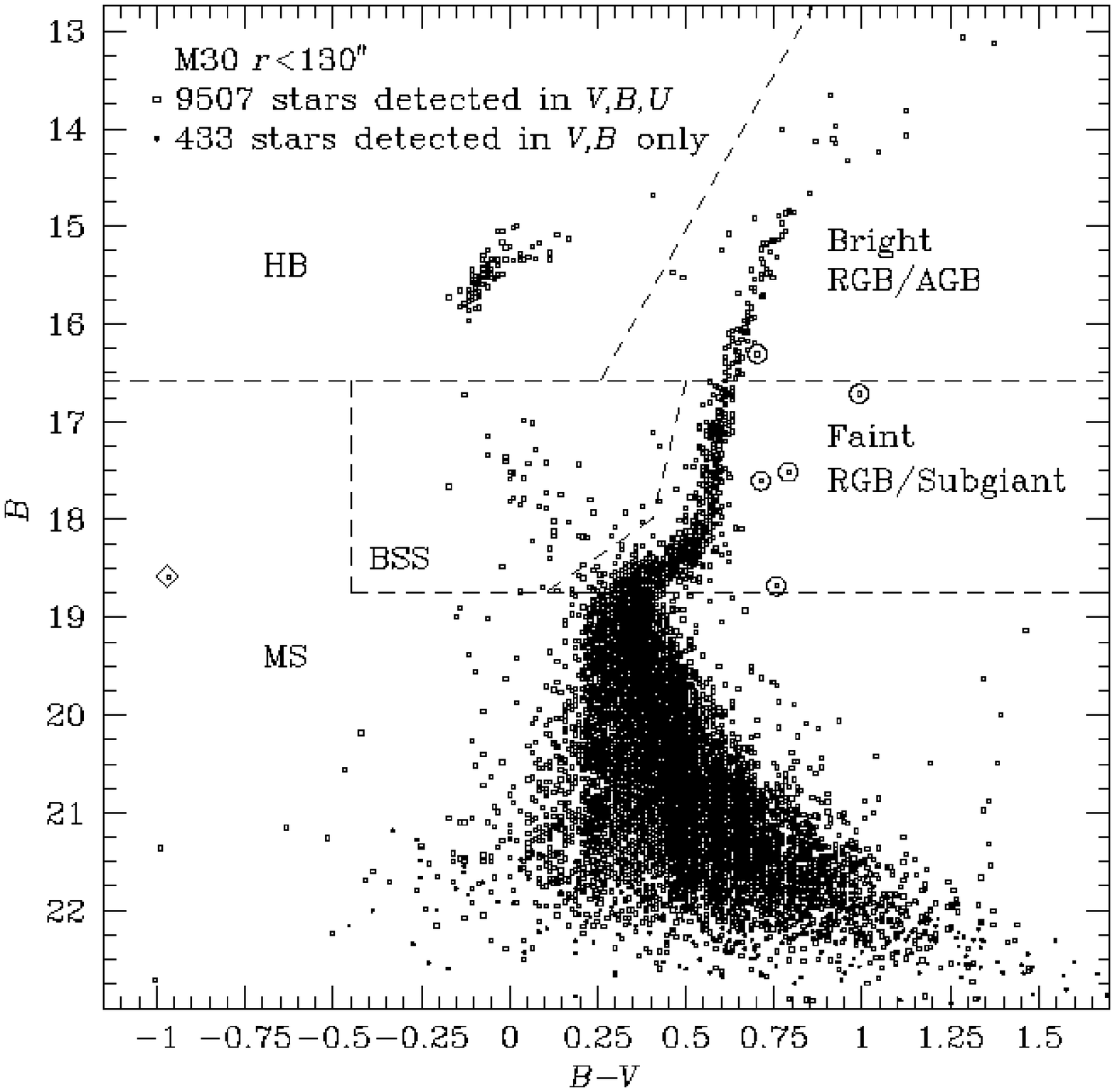}
\caption{A $B$ vs \bv\ color-magnitude diagram of
all 9507~stars in the WFPC2 image of M30 that are detected and
position-matched in the F555W, F439W, and F336W bands bands (squares).  The
433~stars matched only in F555W and F439W are plotted as dots.  The dashed
lines indicated the boundaries that have been used to define various stellar
types.  There are 48~stars in the region labeled BSS.  The large diamond
symbol marks an unusually blue star located close to the cluster center.  The
five~encircled symbols represent stars which are significantly redder than
the red giant branch in all three color-magnitude plots (see
Figs.~\ref{uvbcmd} and \ref{ubucmd}); these may be Galactic field dwarfs in
the foreground. \label{bvbcmd}}
\end{figure}

Figure~\ref{bvbcmd} shows a (\bv,~$B$) CMD of all stars detected
($B\lesssim23$) in the 5.1~arcmin$^2$ area of the WFPC2 mosaic image of M30.
The 9507~stars matched in the $U$, $B$, and $V$ filters are marked as
squares.  An additional 433~stars that are detected and position-matched only
in $B$ and $V$ are marked as dots (most are fainter than $V=21$).  This CMD
is used to assign a stellar type to each star: bright~RGB and asymptotic
giant branch~(AGB) stars, faint~RGB and subgiant stars, HB stars, and blue
straggler stars~(BSSs).  The HB is entirely on the blue side of the RR~Lyrae
instability strip, with its $B$-band brightness becoming fainter towards
bluer $B-V$ colors.  The bright part of M30's RGB and AGB are relatively blue
(relative to the average Galactic globular cluster), in keeping with its low
metallicity: $\rm[Fe/H]=-2.13$ (\cite{metll}).  The MS is reasonably well
defined up to~$\sim2$~mag below the MSTO.  There is a prominent,
well-populated blue straggler sequence in the diagram (Sec.~\ref{BSS}).  The
definition of BSS is somewhat arbitrary, as is the boundary between faint and
bright RGB stars.  The bright vs faint~RGB boundary has been drawn at
$B=16.6$, fainter than the level of the HB (the customary demarcation line),
in order to ensure that the bright~RGB/AGB sample is large enough for a study
of radial gradients.  Note, a similar CMD (based on PC1 data alone) is
presented in \cite{letter}.  Even though the $x$-axis of Fig.~4 of
\cite{letter} is mistakenly labeled as ``$U-V$'', the quantity that is
plotted in that figure is \bv\ color with an arbitrary offset.

\begin{figure}
\plotone{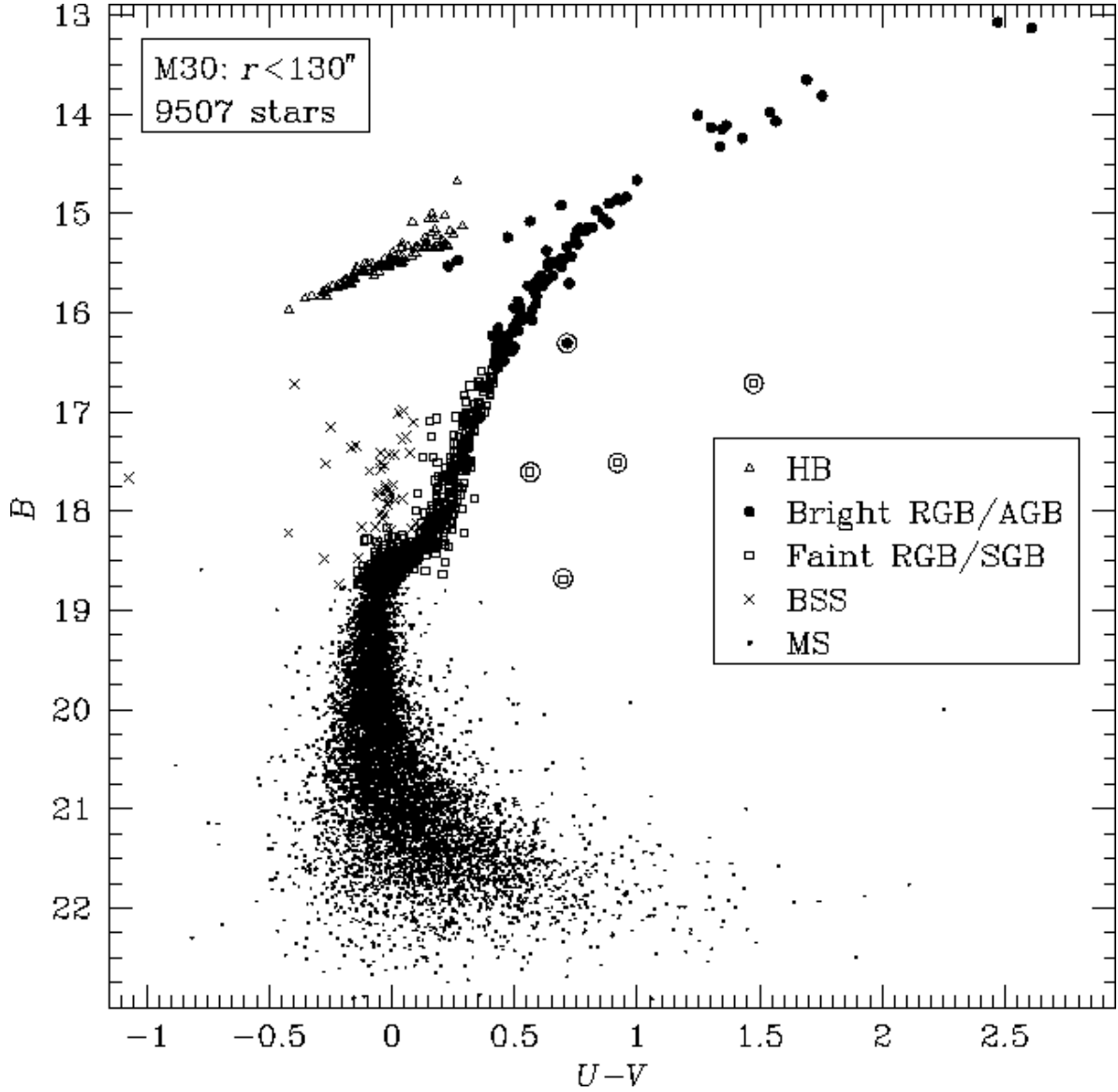}
\caption{Same as Fig.~\ref{bvbcmd} except $B$
magnitude is plotted against $U-V$ color for the 9507~stars matched in all
three~bands.  Different symbols are used to plot the different stellar types,
defined according to the ($B-V$,~$B$) CMD (Fig.~\ref{bvbcmd}).  This
projection distinguishes clearly between the HB, bright RGB, and AGB with the
help of the wide $U-V$ baseline which is sensitive to subtle differences in
stellar temperature.  \label{uvbcmd}}
\end{figure}

\begin{figure}
\plotone{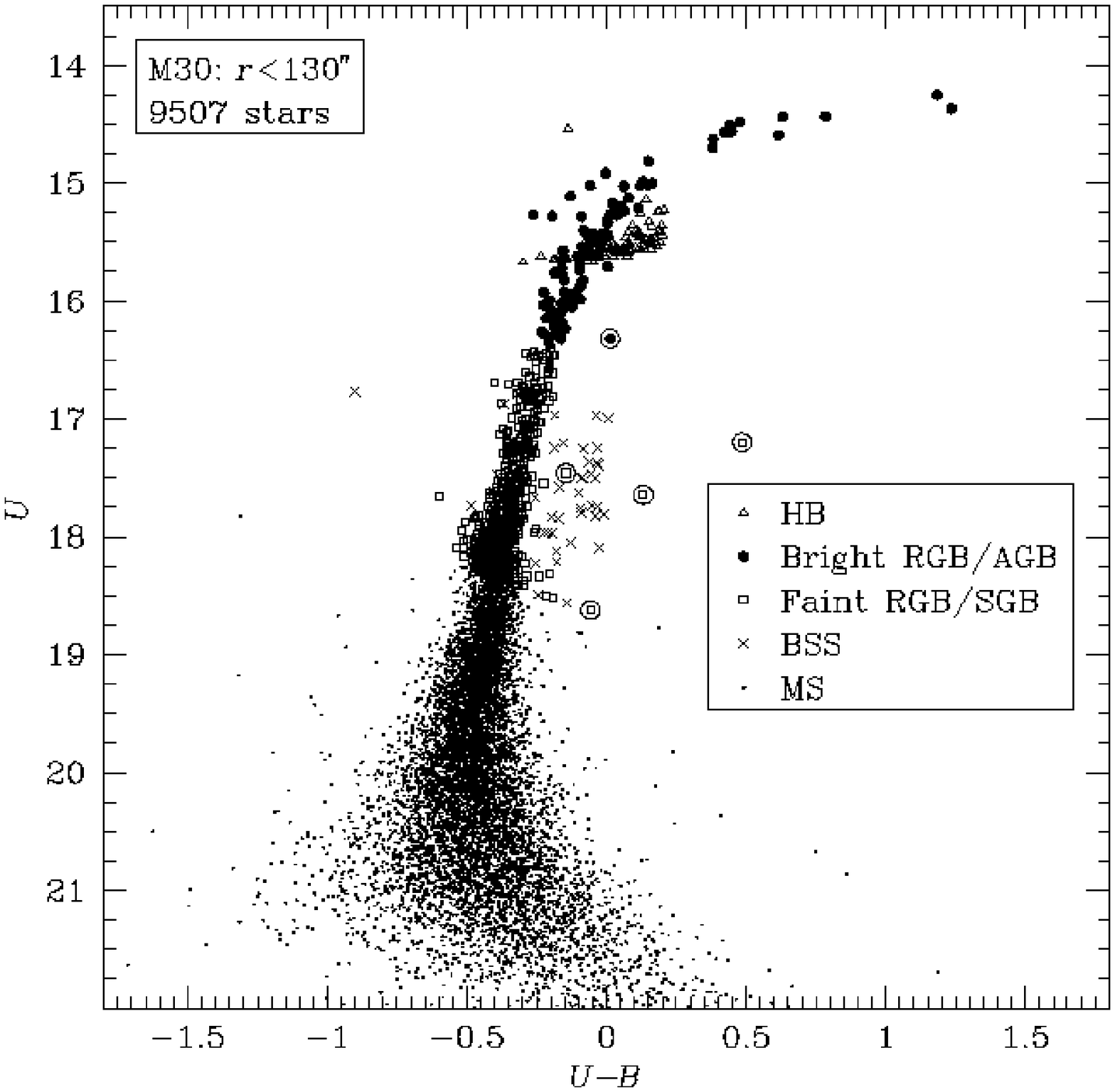}
\caption{Same as Fig.~\ref{uvbcmd} except $U$
magnitude is plotted against $U-B$ color.  The HB is nearly horizontal in
this projection.  The red leak in the WFPC2 F336W filter causes the RGB and
MS to be artificially shifted to the left, so that they overlap with the HB
and BSS sequence, respectively. \label{ubucmd}}
\end{figure}

Using different symbols for different stellar types [as defined by the
(\bv,~$B$) CMD], we plot the same 9507~stars in the ($U-V$,~$B$) and
(\ub,~$U$) CMDs in Figs.~\ref{uvbcmd} and~\ref{ubucmd}, respectively.  The
long $U-V$ color baseline is sensitive to subtle differences in stellar
temperature (despite the red leak in the {\it HST\/} F336W filter), resulting
in a clear separation among the HB, bright RGB, and AGB.  The MS and BSS
sequence are mostly vertical in the $(U-V,~B)$ projection making it difficult
to distinguish between ``true'' BSSs and artifacts caused by the blending of
MSTO stars.  The HB stars all have about the same $U$ brightness, the
bolometric correction to the $U$ magnitude being nearly independent of
temperature for these hot stars, so that the HB feature is nearly horizontal
in the ($U-B$,~$U$) CMD.  This projection highlights the seriousness of the
red leak in the F336W filter which causes the bright RGB and AGB to shift to
the left so that they intersect the HB.  The red leak also causes the
faint~RGB and MS stars to be mapped all the way across to the left (blue)
side of the BSS sequence.  Even with the red leak however, stars of a given
type, as defined in the ($B-V$,~$B$) CMD, tend to remain grouped together in
the ($U-B$,~$U$) projection.  

Note the very blue star in the (\bv,~$B$) CMD: $B-V\sim-0.97$,~$B\sim18.6$
(large diamond symbol in Fig.~\ref{bvbcmd}; star ID\#~2931 in
Table~\ref{datatable}).  This star may possibly be a cataclysmic variable
star based on its location in the CMD, although we are unable to comment on
its variability from our data set alone.  It is located $1\farcs2$~SSW of the
cluster center, well outside the $r=0\farcs6$ (formal) $3\sigma$ error circle
for the cluster centroid, offset by $\Delta\alpha=-2\farcs88$ and
$\Delta\delta=-8\farcs59$ from the reference star (see Sec.~\ref{astrom} for
details on astrometry); this unusually blue star is marked by a mauve circle
in Fig.~\ref{pictm30}.  The apparent proximity of the object to the cluster
center makes it particularly interesting, but the extreme crowding in this
region makes spectroscopic followup a difficult prospect (the star has a
$\Delta{V}=5$~mag brighter RGB neighbor only $0\farcs4$ away).

There are five~relatively bright ($V\sim15.5\>$--$\>18$) objects in the CMDs
which are significantly redder in $U-B$ and $B-V$ (by up to +1~mag) than
typical RGB, subgiant, and MSTO stars of comparable apparent brightness in
M30; these objects have stellar ID\#s of 890, 4735, 9200, 9350, and 9508 in
Table~\ref{datatable} and are indicated by encircled symbols in
Figs.~\ref{bvbcmd}$\>$--$\>$\ref{ubucmd}.  One cannot rule out the
possibility that these are cluster stars with abnormal spectral
characteristics.  However, the fact that these five~objects are {\it not\/}
concentrated towards M30's center casts serious doubt on their cluster
membership: four~out of the~five lie beyond $r=1\arcmin$, whereas only about
15\% of cluster members within the WFPC2 image are expected to lie beyond
this radius.  No radial velocity measurements are available for these
five~red objects (K.~Gebhardt, private communication); M30's large negative
radial velocity, $v_{\rm M30}\sim-187$~km~s$^{-1}$ (\cite{zaggia92}), should
make it easy in principle to distinguish field stars from cluster members.
On the basis of a realistic model of the Galaxy, \cite{bahrat} predict a
surface density of field star interlopers of about 3~arcmin$^{-2}$ to
$V\sim23$, or a total of 15~interlopers within the area of the WFPC2 mosaic.
Most of these field stars are expected to be faint: less than four~field
stars are expected to be brighter than  $V\sim21$ and less than one~star
(0.37) is expected to be brighter than $V=19$.  The fact that even field
stars tend to cluster together implies that the probability of finding
five~bright field stars could be higher than the Poisson estimate.  If these
five~red objects in the the M30 WFPC2 data set are foreground MS stars of
similar metallicity, there is a 3~mag spread in distance modulus (factor of
4~in distance) amongst them.  These objects are even less likely to be
background galaxies.  Field stars are expected to outnumber field galaxies
for $V<20$, even at high Galactic latitude.  Moreover, our PSF fitting
procedure is designed to identify and eliminate extended objects (galaxies);
no such objects are found in the M30 data set.

\begin{figure}
\plotone{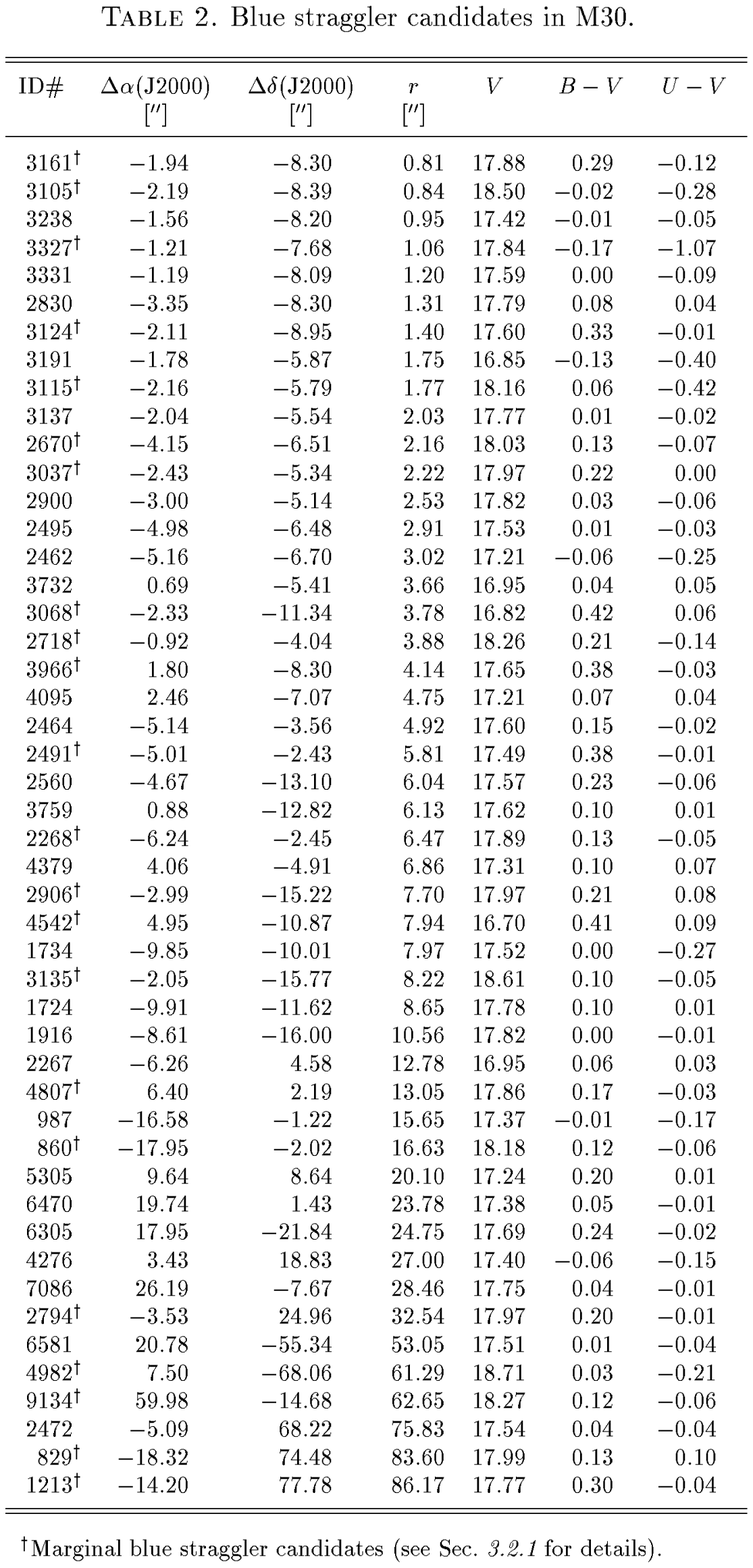}
\end{figure}
\placetable{bsstable}
\begin{table}\dummytable\label{bsstable}\end{table}

\subsection{Blue Stragglers}\label{BSS}
\subsubsection{Identification}\label{bssident}
The ($B-V$,~$B$) CMD of M30 presented in Fig.~\ref{bvbcmd} shows a large
collection of BSS candidates in the region between the MSTO and the extended
blue tail of the HB.  This sequence of BSS candidates lies roughly where the
continuation of the MS would be, if it were to be extrapolated beyond
(i.e.,~brighter and bluer than) the MSTO.  These stars are, for the most
part, distinct from HB and MSTO stars in all three color-magnitude
projections (see Figs.~\ref{bvbcmd}$\>$--$\>$\ref{ubucmd}).

Table~\ref{bsstable} lists the positions, brightnesses, and colors of the
48~BSS candidates outlined by the dashed lines in Fig.~\ref{bvbcmd}.  All
positions are on the equinox J2000 coordinate system and are measured with
respect to the reference star as described in Sec.~\ref{astrom}.  The stars
in Table~\ref{bsstable} are arranged in order of increasing projected
distance from the cluster center.  The stellar ID\#s for the BSS candidates
are the same as those in the complete (electronic) version of
Table~\ref{datatable}.

A ``$\dag$'' symbol is used to mark marginal candidates in
Table~\ref{bsstable}, those for which the BSS designation may be in some
doubt.  Results from earlier image simulations (Papers~IV and V) indicate
that blending and/or errors in the photometry of subgiant/MSTO stars can
contaminate the BSS portion of the CMD.  We have been very conservative in
designating ``marginal'' BSS candidates: these include all the ones fainter
than $B=18$ (since this region may be populated by blend artifacts), those
that are close to the edge of the selection region, and outliers that are
separated from the upward extrapolation of the MS track.  It is possible that
some fraction of the marginal BSS candidates in M30 are actually
subgiant/MSTO stars that have been scattered away from their fiducial
locations in the CMD due to measurement error.  Contamination is likely to be
most severe at the base of the BSS sequence where it meets the subgiant
branch and MSTO ($B\approx18.5$, $B-V\approx0.25$), so this region has
deliberately been excluded from the BSS defining box from the outset
(Fig.~\ref{bvbcmd}).  Our BSS sample is likely to be incomplete at the faint
end (near the MSTO), as the defining box is liable to exclude BSS if their
luminosity $L_{\rm BSS}\lesssim1.4L_{\rm MSTO}$.  The exact balance between
the competing effects of BSS sample contamination and incompleteness depends
on the details of the somewhat arbitrary BSS defining criteria used.  The M30
BSS population is unlikely to be contaminated by field star interlopers; less
than one~field star is expected in the region of the (\bv,~$B$) CMD brighter
and bluer than the MSTO based on the Bahcall-Soneira model of the Galaxy
(\cite{bahcson}; Ratnatunga~\& Bahcall 1985).
 
\subsubsection{Specific Frequency}\label{fbss}

The specific frequency of BSSs in M30 is calculated according to the
definition of \cite{bolte93}: $F_{\rm BSS}=N({\rm BSS})/N(V<V_{\rm HB}+2)$.
Although M30's HB stars do not all have the same $V$ brightness
(Fig.~\ref{bvv}), the HB has a well defined brightness of $V\approx15.1$ at
the location of the RR~Lyrae instability region ($\bv\sim0.3$) before
dropping to fainter $V$ magnitudes at the blue end.  There are
332~RGB/AGB/HB/bright subgiant stars in the $5.1$~arcmin$^2$ WFPC2 mosaic
image ($r\lesssim130\arcsec$) that fulfill the criterion for the normalizing
population and 48~BSSs---an overall specific frequency of $F_{\rm
BSS}=0.14\pm0.02$ ($1\sigma$ Poisson error).  This statistic is computed as a
function of radius and the results are listed in Table~\ref{spfreq}.  The
numbers listed in Table~\ref{spfreq} are slightly different from those quoted
in \cite{letter}; this is due in part to the inclusion of data from the WF
CCDs and in part to a slightly revised defining criterion for BSS candidates.

\vskip .2in
\centerline{{\sc Table 3.} Blue straggler and subgiant/turnoff specific frequency vs
radius in M30.}
\vskip .15in

\begin{minipage}{5.5in}
\begin{tabular}{rccccc}
\hline\hline
{$r_{\rm lim}$}	& {$N_{\rm BSS}$} &
{$N(V<V_{\rm HB}+2)$} & {$F_{\rm BSS}$} & {$N_{\rm SGTO}$} & {$N(V<V_{\rm HB}+2)/N_{\rm SGTO}$} \\
$[\arcsec]$ & {} &  {} &  {} & {} & {}\\ \hline

10	& 31	& ~$\>$79 &	0.392 	&	244	& 0.324	\\
20	& 36	& 144 	& 	0.250 	&	429	& 0.336	\\
30	& 41	& 192 	& 	0.214 	& 	528	& 0.364 \\
60	& 43	& 267 	& 	0.161 	& 	750	& 0.356 \\
100	& 48	& 327 	& 	0.147 	& 	921	& 0.355 \\ \hline\hline
\end{tabular}
\end{minipage}
\vskip .1in

\placetable{spfreq}
\begin{table}\dummytable\label{spfreq}\end{table}

The BSS specific frequency found in the inner $20\arcsec$ of M30, $F_{\rm
BSS}(r<20'')=0.25\pm0.05$, is higher than the specific frequency found in
other well studied (e.g.,~at {\it HST\/} resolution) globular clusters,
including ones whose density is comparable to or greater than that of M30,
such as M15 (cf.~\cite{sosin}; Paper~VI).  In fact, the {\it overall\/} BSS
frequency in the M30 WFPC2 data set, $F_{\rm BSS}=0.14$, falls at the high
end of the typical factor of~3 range seen in Galactic globular clusters.

If the marginal BSS candidates in M30 are excluded from the analysis, the
remaining (secure) BSS candidates have a specific frequency of $F_{\rm
BSS}=0.08$ in the overall data set and $F_{\rm BSS}=0.13$ in the central
$r<20''$.  These values should be treated as lower bounds to the ``true'' BSS
frequency in M30---a fair fraction of the marginal candidates (possibly the
majority) are likely to be real BSSs, and the defining box is liable to have
missed faint BSSs (Sec.~{\it\ref{bssident}}).  For the purpose of comparing
M30's BSS frequency to that of other clusters, it is best to consider $F_{\rm
BSS}$ values based on the full set of BSS candidates (secure+marginal)
because: (1)~estimates of $F_{\rm BSS}$ in other clusters tend to include
marginal candidates, and (2)~the BSS defining criteria used in this paper
(Fig.~\ref{bvbcmd}) are similar to those used in other studies.

\subsubsection{Central Concentration}\label{bsscconc}

The value of $F_{\rm BSS}$ in M30 displays a dramatic increase towards the
cluster center (see Table~\ref{spfreq} and Sec.~{\it\ref{center}}).  The BSS
frequency in the $20\arcsec<r<130\arcsec$ region of the WFPC2 mosaic image
is~$0.06\pm0.02$, a factor of~4 lower than in the inner $r<20\arcsec$
($F_{\rm BSS}=0.25$), but in keeping with the value seen in other clusters
(Paper~VI).  The concentration of the M30 BSSs towards the cluster center is
consistent with the idea that they are more massive than individual
RGB/subgiant stars (either merger remnants or members of close binary
systems), and have consequently sunk deeper into the cluster potential well
(\cite{king91}).

Restricting the M30 BSS sample to only secure candidates (i.e.,~those {\it
not\/} marked by a ``$\dag$'' in Table~\ref{bsstable}), the specific
frequency is seen to increase by a factor of~3 from $F_{\rm BSS}=0.04$ in the
outer $20''<r<130''$ region to $F_{\rm BSS}=0.13$ in the central $r<20''$
region.  The sample of secure BSS candidates, being well removed from the
subgiant/MSTO region of the CMD, is uncontaminated by blends.  This is firm
proof that M30's BSSs are indeed physically concentrated towards the cluster
center; the observed concentration is {\it not\/} an artifact of increased
photometric scatter/blending in the crowded central region of the cluster.
In fact, the data in Tables~\ref{bsstable} and \ref{spfreq} suggest that the
majority of even the marginal candidates in M30 are faint (but real) BSSs
which are centrally concentrated in a similar manner to their brighter (and
hence more secure) counterparts.

We investigate whether the high BSS fraction in M30's inner region is
produced by anomalies in the normalizing population of RGB/AGB/HB/bright
subgiant stars ($V<V_{\rm HB}+2$) by comparing the radial distribution of
this normalizing population to that of faint subgiant/turnoff (SGTO) stars in
the cluster.  The SGTO class is defined by the criterion: $V_{\rm
HB}+2<V<V_{\rm MSTO}$, where $V_{\rm MSTO}=18.6$.  The fraction, $F_{\rm
SGTO}\equiv{N}(V<V_{\rm HB}+2)/N(V_{\rm HB}+2<V<V_{\rm MSTO})$, is consistent
with being constant throughout the central $r<2\arcmin$ of M30
(Table~\ref{spfreq}); the apparent 5\%--10\% decrease in $F_{\rm SGTO}$ in
the inner $20\arcsec$ compared to the global value is not statistically
significant ($\lesssim1\sigma$ effect), and this has a negligible effect on
the measurement of $F_{\rm BSS}$.  Furthermore, a study of the LF of M30's
evolved stellar population shows that the abundance of RGB/AGB stars relative
to fainter stars (faint subgiant, MSTO, and MS stars) is not anomalously low;
rather, their abundance is a little higher than normal (see
Sec.~\ref{lf_vs_model}).   

\subsubsection{Blue Straggler Luminosity Function}

\begin{figure}
\plotone{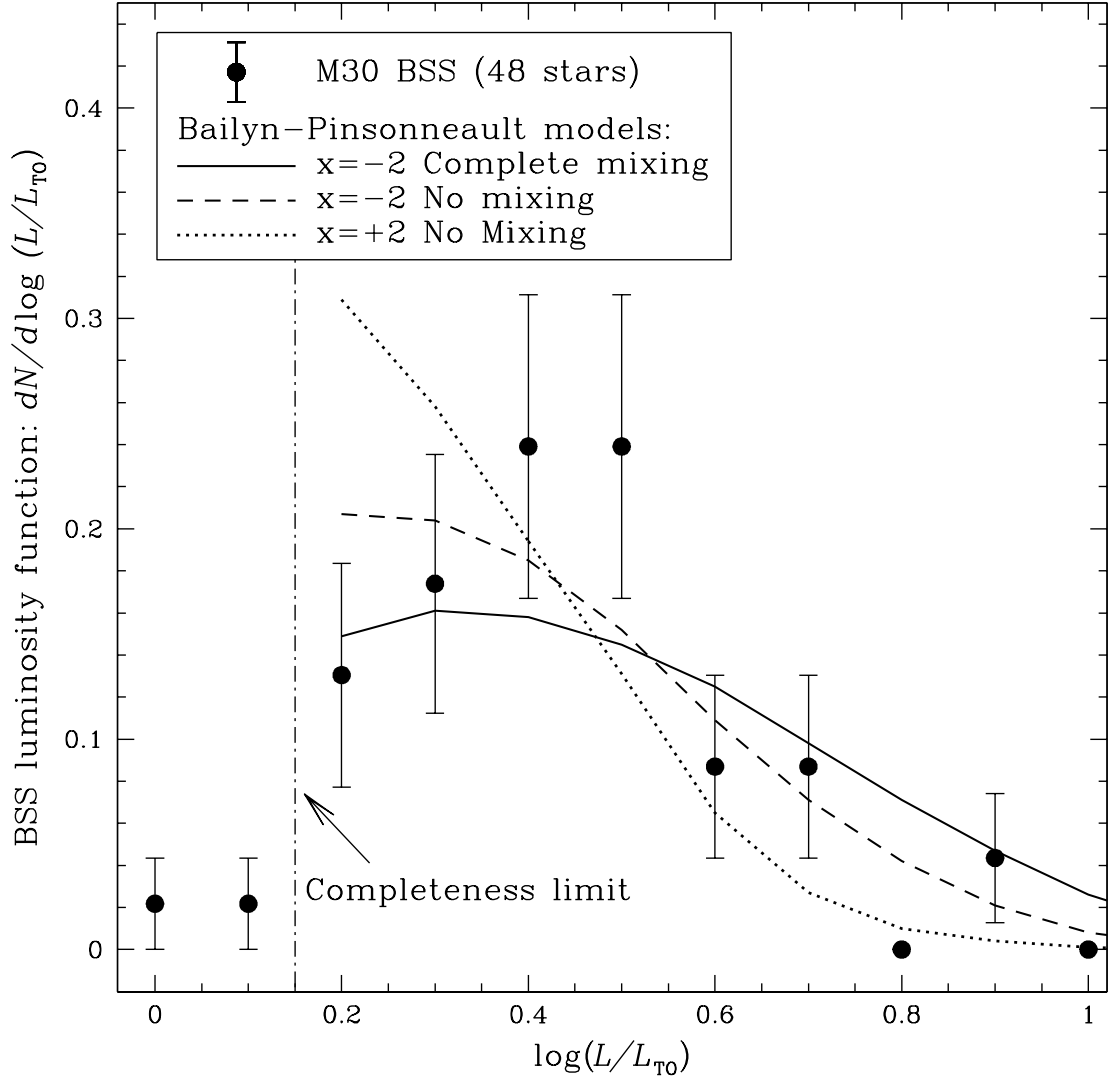}
\caption{Luminosity function of the blue straggler
stars in M30 (bold dots with Poisson error bars).  The BSS sample is likely
to be incomplete fainter than $\log(L/L_{\rm MSTO})=0.15$ (dot-dashed
vertical line) due to our selection criterion.  Three~theoretical LF models
(Bailyn \& Pinsonneault 1995) are plotted for comparison, reflecting a
variety of BSS formation scenarios: $x=-2$ vs $x=+2$ input mass function
slopes (appropriate for stellar collisions and binary mergers, respectively),
and complete mixing vs no mixing for the chemical composition of the newborn
BSS.  The models are based on a metal abundance of $Z=6.25\times10^{-4}$
(${\rm[Fe/H]}=-1.7$) and, like the data, are normalized to unity over the
range $\log(L/L_{\rm MSTO})\geq0.15$.  \label{bslum}}
\end{figure}

Theoretical studies of the formation of BSSs indicate that the BSS LF may be
a good discriminant between different formation scenarios (\cite{bsstheory}).
For example, stellar collisions produce a tail of very luminous BSSs
while mergers of primordial binaries do not.  Bailyn~\& Pinsonneault adopt
a top heavy input mass function for stellar collisions since these are
important in the dense central regions of clusters where massive stars are
expected to be concentrated as a result of mass segregation; input masses for
the primoridal binary merger case are likely to be drawn from a more normal
(e.g.,~\cite{salpeter}; $x=+1.35$) mass function.  The authors also consider
two~extreme options for the compositional structure of a newborn BSS that
ought to bracket the realistic range of possibilities---they dub these
`collisional' chemistry, which corresponds to complete mixing yielding a
chemically homogenous product, and `binary merger' chemistry, which
corresponds to no mixing.

Figure~\ref{bslum} shows the observed BSS LF in M30 (bold dots with Poisson
error bars) compared to various Bailyn~\& Pinsonneault model BSS LFs for
metal poor stars: $Z=6.25\times10^{-4}$ or $\rm[Fe/H]=-1.7$.  The BSS
luminosities are estimated from their $U$ brightnesses, with a $+0.25$~mag
correction to the observed $U_{\rm MSTO}=18.5$ to account for red leak and an
additional $-0.1$~mag differential bolometric correction.  We expect our
sample of M30 BSSs to be incomplete for $L_{\rm BSS}/L_{\rm MSTO}<1.4$ or
${\rm log}(L_{\rm BSS}/L_{\rm MSTO})<0.15$ (indicated by vertical dot-dashed
line in Fig.~\ref{bslum}); this region is too close to the MSTO and our BSS
selection boundary has been designed to avoid contamination from MSTO and
subgiant stars.  The model LFs and the M30 data are normalized to unity over
the ``complete'' range, $\log(L/L_{\rm MSTO})\geq0.15$.

The shape of the LF of BSSs in M30 bears a general resemblance to the shape
of the collisional model LFs ($x=-2$).  The peak in the M30 LF at ${\rm
log}(L_{\rm BSS}/L_{\rm MSTO})\sim0.5$ and the suggestion of a tail to higher
BSS luminosities are not reproduced in the primordial binary merger model LF:
An $x=+2$ input mass function binary merger model is shown in
Fig.~\ref{bslum}; in fact, all merger models with $x\gtrsim1$ have a
qualitatively similar shape to the $x=+2$ case (\cite{bsstheory}).
The intermediate luminosity peak in M30's BSS LF is also apparent from the
($B-V$,~$B$) CMD in Fig.~\ref{bvbcmd}: note the concentration of BSSs and the
hint of a ``gap'' between the BSS sequence and MSTO.  While BSSs are more
abundant in M30's central region than in any of the other clusters studied to
date, the total number of stragglers found in the WFPC2 image of M30 is too
small to draw any firm conclusions about the shape of their LF.  It is fair
to say though that the data are strongly suggestive of a collisional origin
for the BSSs.

\subsection{Population Gradients}

In order to study radial variations in the mix of M30's stellar populations,
the stars detected in the full WFPC2 mosaic are divided into eight~annuli
around the cluster center, as in Sec.~\ref{acccomp}:
(1)~$r<5\farcs00$, 
(2)~$5\farcs00\leq{r}<9\farcs80$,
(3)~$9\farcs80\leq{r}<15\farcs41$,
(4)~$15\farcs41\leq{r}<23\farcs2$,
(5)~$23\farcs2\leq{r}<35\farcs8$,
(6)~$35\farcs8\leq{r}<51\arcsec$,
(7)~$51\arcsec\leq{r}<71\arcsec$, and
(8)~$71\arcsec\leq{r}<130\arcsec$.
The limiting radii of the eight~radial bins have been chosen so that each bin
contains approximately the same number of stars with $V<18.6$, a limiting
magnitude for which the samples at all radii are expected to be complete.  

\begin{figure}
\plotone{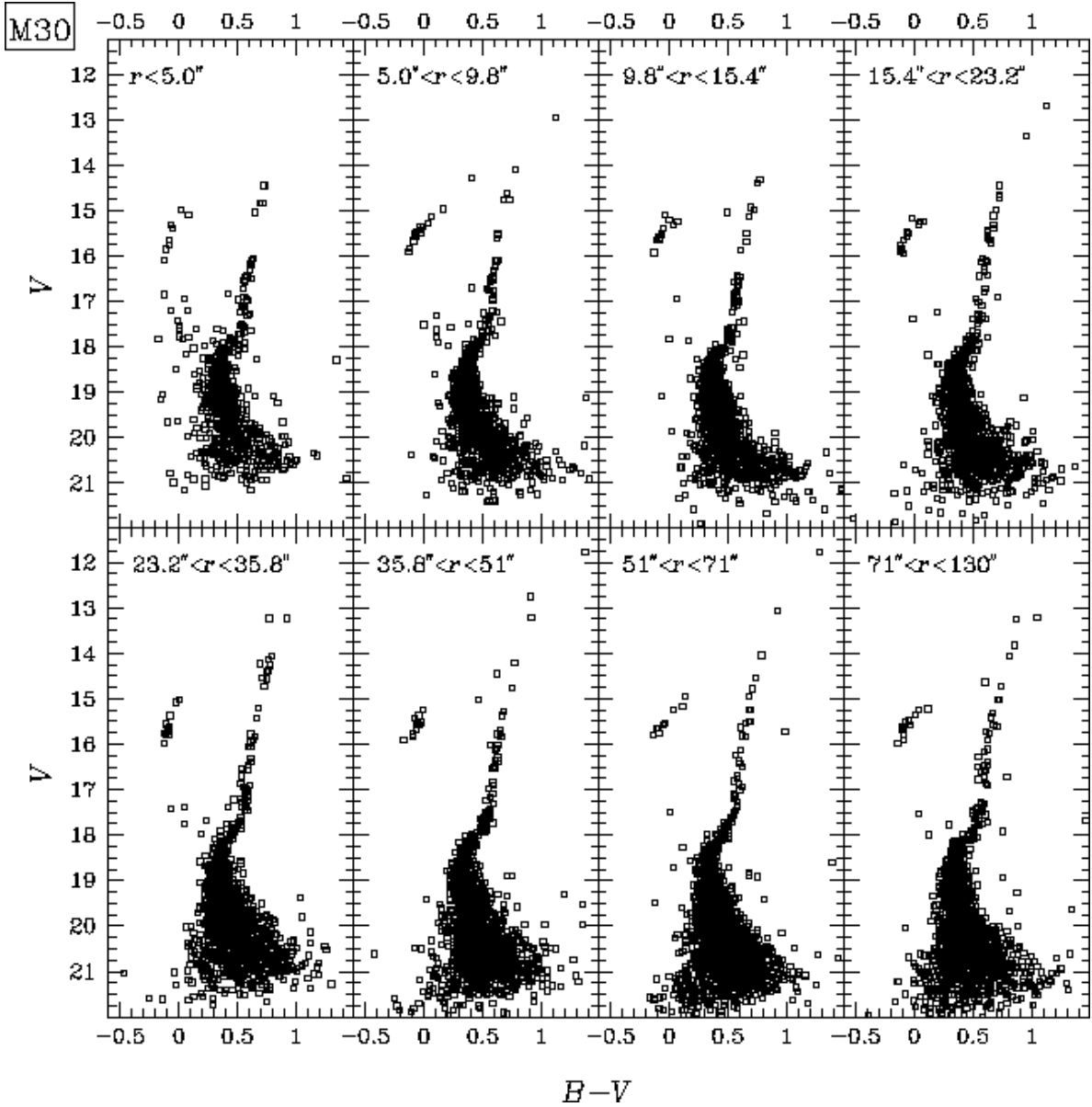}
\caption{A composite of $V$ vs $B-V$
color-magnitude diagrams in eight~radial bins around the center of M30.  The
limiting radii have been chosen to ensure that the bins contain equal numbers
of stars with $V<18.6$.  The outer five~bins (\#4--\#8) represent partial
annuli.  The innermost bin contains an excess of blue stragglers and appears
to be deficient in bright RGB stars relative to the bins farther out.  The
photometric accuracy improves and degree of completeness for faint stars
increases with increasing radius.  \label{bvv}}
\end{figure}

\subsubsection{Comparing the Radial Distribution of Various Stellar Types}
\label{center}

In this section, we compare the radial density distributions of the various
types of evolved stars found in M30's central region.  Figure~\ref{bvv} is a
composite of the ($B-V$,~$V$) CMDs in each of the eight~radial bins.  The
morphology of the stellar distribution in this CMD is qualitatively similar
to that in the ($B-V$,~$B$) CMD (Fig.~\ref{bvbcmd}), except that both the RGB
and HB are slightly more extended and vertical than in the latter CMD.  There
are no striking differences in the mix of stellar types from one annulus to
another, aside from the obvious excess of BSSs in the inner bins (see also
Sec.~{\it\ref{bsscconc}\/} and Table~\ref{spfreq}).  There is an indication
that the central regions are deficient in bright RGB/AGB stars
($V\lesssim16$); the innermost bin also appears to be slightly deficient in
HB stars (the degree and significance of these effects are quantified below).
The scatter due to photometric error decreases, and the degree of
completeness for stars fainter than about $V=20$ increases, as one moves from
the dense central region to the less crowded cluster outskirts.
\vskip .2in
\centerline{{\sc Table 4.} Relative radial distribution of evolved
stellar populations in M30.}
\vskip .15in
\begin{minipage}{6.8in}
\begin{tabular}{l|r|r|r|r|r|r|r|r|r}
\hline\hline
{} & \multicolumn{8}{c}{Radial Bins\footnote{\begin{tabular}{llll}
(1)~$r\leq5\farcs00$, & 
(2)~$5\farcs00<r\leq9\farcs80$, &
(3)~$9\farcs80<r\leq15\farcs41$,&
(4)~$15\farcs41<r\leq23\farcs2$,\\ 
(5)~$23\farcs2<r\leq35\farcs8$,&
(6)~$35\farcs8<r\leq51\arcsec$,&
(7)~$51\arcsec<r\leq71\arcsec$,&
(8)~$71\arcsec<r<130\arcsec$.\end{tabular}}} &
{} \\ \hline
{Stellar~Subsample}	& 
{(1)} & {(2)} & {(3)} & {(4)} & 
{(5)} & {(6)} & {(7)} & {(8)} &
{$\overline{n}\pm1\sigma$}\\ \hline
Post MSTO stars ($V<18.6$) & 160 & 159 & 155 & 159 & 157 & 157 &153 & 160 &
$158\pm12.5$\\
Bright stars ($V<V_{\rm HB}+2$) & 36 & 43 & 40 & 43 & 47 & 40 & 36 & 47 &
$41.5\pm6.44$\\
All stars ($V\la23$) & 645 & 915 & 939 & 1099 & 1278 & 1449 & 1647& 1535
&$1188\pm34.5$\\
Faint RGB/Subgiants & 93 & 86 & 87 & 85 & 87 & 104 & 96 & 85 &
$90\pm9.51$\\
Bright RGB/AGB & 5  & 8  & 9  & 16  & 16 & 13  & 15   & 16 & $12.3\pm3.50$\\
HB & 8  & 16 & 13 & 12 & 13 & 11  & 9   & 16 & $12.3\pm3.50$\\
BSS & 21 & 10 & 3 & 3 & 5 & 0 & 3 & 3 & $6.0\pm2.45$\\
\hline\hline
\end{tabular}
\end{minipage}
\vskip .1in
\placetable{binned}
\begin{table}\dummytable{\label{binned}}\end{table}

Table~\ref{binned} lists the number of stars of each type in the eight~radial
bins.  The division of stars into various stellar types is defined by the
dashed lines in Fig.~\ref{bvbcmd}.  The {\it total\/} number of stars in each
radial bin increases towards larger radii due to the increased probability of
detecting faint MS stars ($V\gtrsim20$) in the relatively sparse outer parts.
Faint~RGB stars and subgiants make up the bulk of the complete reference
sample used to define the limiting radii, so it is not surprising that their
numbers are roughly constant across the bins.  There is a $2.1\sigma$
depletion of bright~RGB/AGB stars in the inner bin compared to the average of
all bins.  The inner depletion is slightly more significant ($2.6\sigma$)
relative to the average of the outer four~bins ($15.0\pm3.9$) rather than the
overall average; the overall average is biased low due to the RGB depletion
in the center.  The bright RGB sample size is too small to tell whether the
central deficiency is stronger for the brightest giants (those with $V<V_{\rm
HB}$ or $V\lesssim15$) than for the full set of bright RGB stars
($V\lesssim16$ as defined in Sec.~\ref{cmds}); the effect is less {\it
significant\/} for the $V\lesssim15$ subset than for the full set, but this
is due to the increased Poisson error associated with the brighter subset.
The slight apparent depletion of HB stars in the central 5\arcsec\ compared
to the overall average is not statistically significant ($1.2\sigma$
effect).

\begin{figure}
\plotone{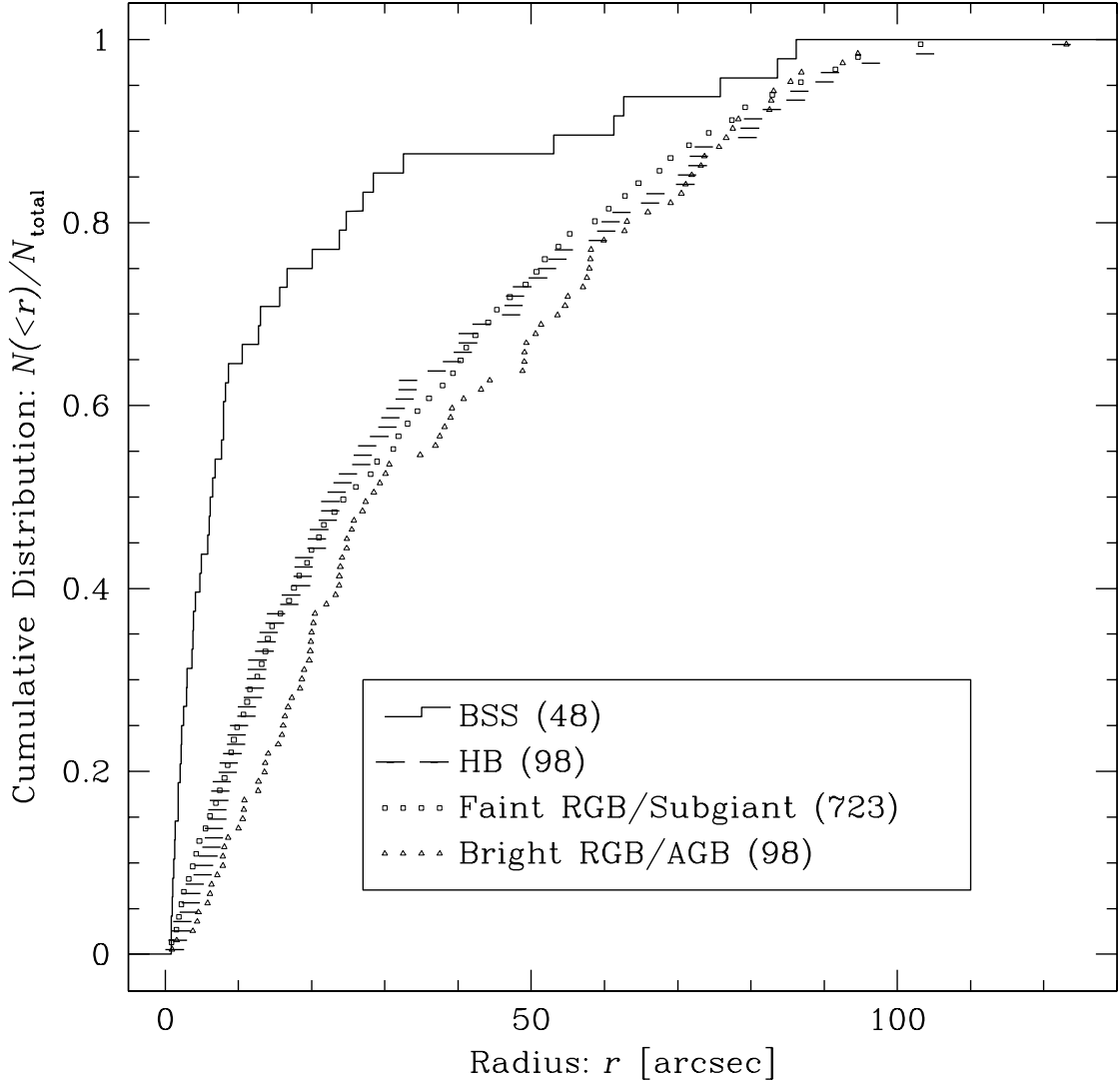}
\caption{The cumulative radial distribution of
various types of post main sequence stars in the WFPC2 image of M30
($r<130''$).  Only every tenth faint RGB/subgiant star is shown for the sake
of clarity.  The numbers within parentheses indicate the number of stars of
each type.  The BSSs are significantly more centrally concentrated and the
bright~RGB/AGB stars appear to be somewhat less centrally concentrated than
the faint RGB/subgiant normalizing population.  \label{cumu}}
\end{figure}

\begin{figure}
\plotone{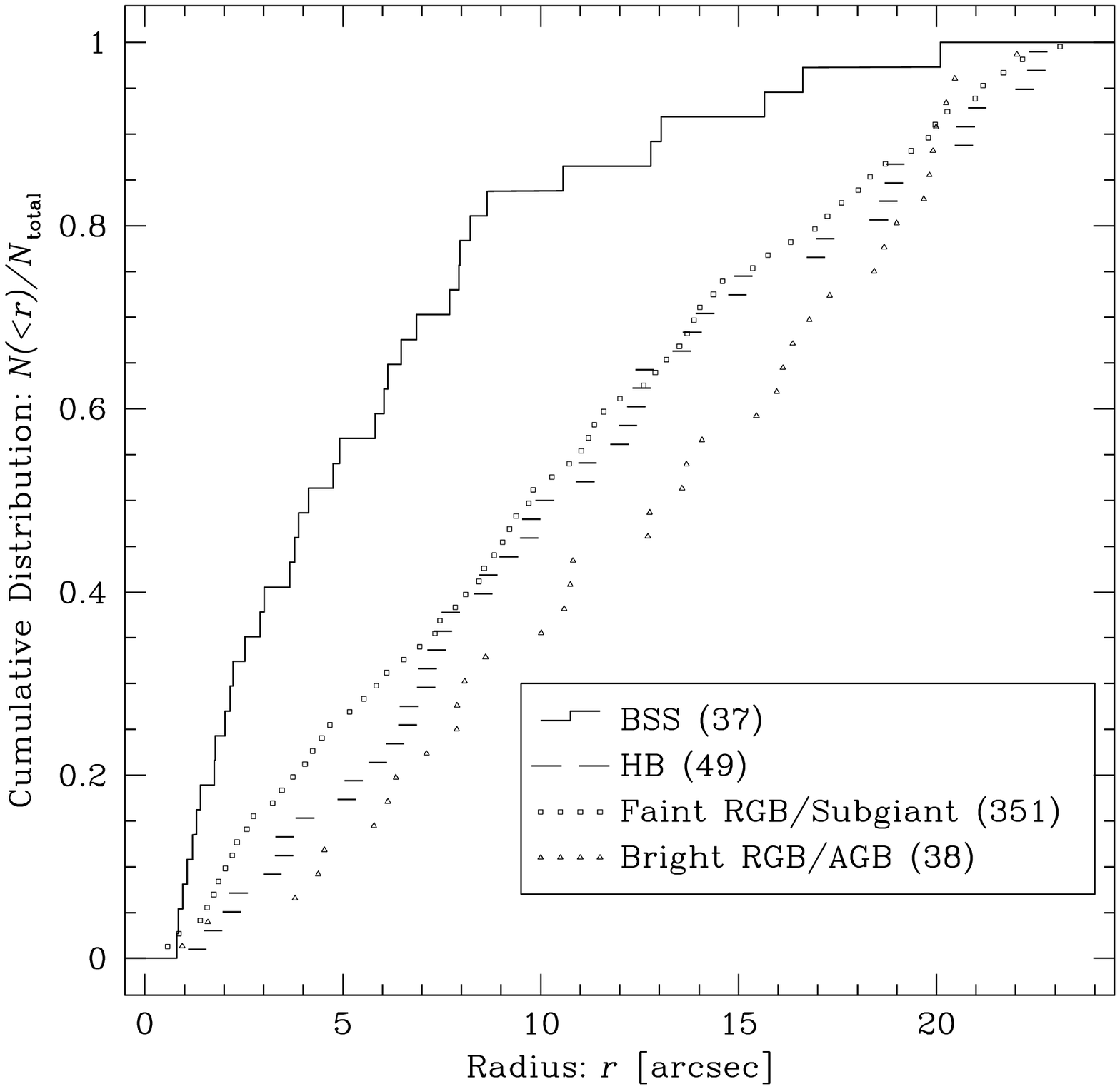}
\caption{Same as Fig.~\ref{cumu} for the inner
$r<23''$ of M30.  Only every fifth faint~RGB/subgiant star is shown for
clarity.  The trends observed in the full sample (Fig.~\ref{cumu}) persist in
the inner region of M30---both the strong central concentration of BSSs as
well as the slight central deficiency of bright~RGB/AGB stars.
\label{cumuinner}}
\end{figure}

Figure~\ref{cumu} shows the cumulative radial distributions of different
types of stars in the core of M30 over the full area of the WFPC2 mosaic
($r<130''$).  Figure~\ref{cumuinner} shows the cumulative radial distribution
for the inner $r<25\arcsec$ region of the cluster.  A two-sided
Kolmogorov-Smirnov~(KS) test is used to measure the probability that a given
pair of stellar types is drawn from the same underlying radial distribution
function.  We use the faint~RGB sample as the comparison population for KS
tests as this phase of stellar evolution is relatively uncomplicated and well
understood.  Faint~RGB, bright~RGB/AGB, and HB stars are expected to have the
same radial distribution because:
(1)~stars evolving ``normally'' should proceed from faint~RGB/subgiant to
bright~RGB/AGB to HB without a significant change in mass, and 
(2)~even if the masses of these different stellar types were slightly
different because of mass loss during the bright RGB phase, mass segregation
proceeds on a dynamical timescale which is longer than the evolutionary
timescale in these phases.

The KS probability that the BSSs are drawn from the same radial distribution
as the faint~RGB/subgiant population is less than $10^{-7}$.  The cumulative
radial distribution plots show that the BSSs are strongly centrally
concentrated with respect to all the other stellar types, and that this trend
is strong even within the inner $25''$ of the cluster (Figs.~\ref{cumu} and
\ref{cumuinner}), in keeping with the specific frequency vs radius
calculation in Sec.~{\it\ref{bsscconc}\/} and Table~\ref{spfreq}.  The
bright~RGB/AGB and faint~RGB/subgiant populations have different radial
distributions, with only a 5.4\% KS probability of being drawn from the same
distribution.  In contrast, the radial distributions of the HB and
faint~RGB/subgiant stars are consistent with each other.

Previous studies of M30 have noted that its central region is deficient in
bright~RGB stars relative to its outer region (cf.~\cite{piotto88}).  The
cause of the bright RGB depletion in M30 and other post-core-collapse
clusters remains something of a mystery.  Djorgovski et~al.\ (1991) examine a
variety of physical mechanisms---tidal interactions, stellar collisions,
Roche lobe overflow of an evolving red giant in a tight binary system,
etc.---but find that none of these provides a convincing explanation of the
observations.  The general consensus appears to be that binaries play {\it
some\/} role in causing the central depletion of bright~RGB stars; this has
motivated several detailed, but as yet inconclusive, theoretical
investigations of the effect of binaries on post MS evolution
(cf.~\cite{taamlin}; \cite{dantona}).

\subsubsection{Which Stellar Populations are Responsible for the Radial Color
Gradient Near M30's Center?}
\label{colorgrad}

In order to study the radial color gradient in the integrated starlight of
the cluster, we construct mosaic images from the WFPC2 data in the F439W
($B$) and F555W ($V$) bands (greyscale image in Fig.~\ref{pictm30}) using the
{\sc metric} task in {\sc iraf}/{\sc stsdas} which corrects for
distortion/aberration in each CCD image and inter-CCD rotations and
translations.  Standard aperture photometry routines (with sky subtraction)
are used to measure the integrated flux in each band in each of the
eight~radial bins defined in Sec.~\ref{acccomp}.  The background ``sky''
level is estimated from relatively sparse areas of the mosaic image.  Sky
counts make up as much as 22\% of the total counts in the outermost radial
bin ($71''<r<130''$) of the F439W image; the fractional contribution of sky
counts is lower in the F555W band, and decreases sharply at smaller radii in
both bands.  The integrated cluster fluxes are converted to $B$- and $V$-band
surface brightnesses using the photometric calibration parameters (zeropoint
and color transformation) described in Sec.~\ref{dataproc}.  The
``effective'' radius of each bin is taken to be the median radial distance
from the cluster center of all evolved stars in that bin.  This is more
representative than the simple average or area-weighted average of the
limiting radii of the bin; the distinction is especially important for the
innermost bin which covers the largest dynamic range in radius, and for the
outermost bin whose outer edge is defined by the outline of the WFPC2 mosaic
image and this causes the actual number of stars detected to be a strongly
decreasing function of radius.

\begin{figure}
\plotone{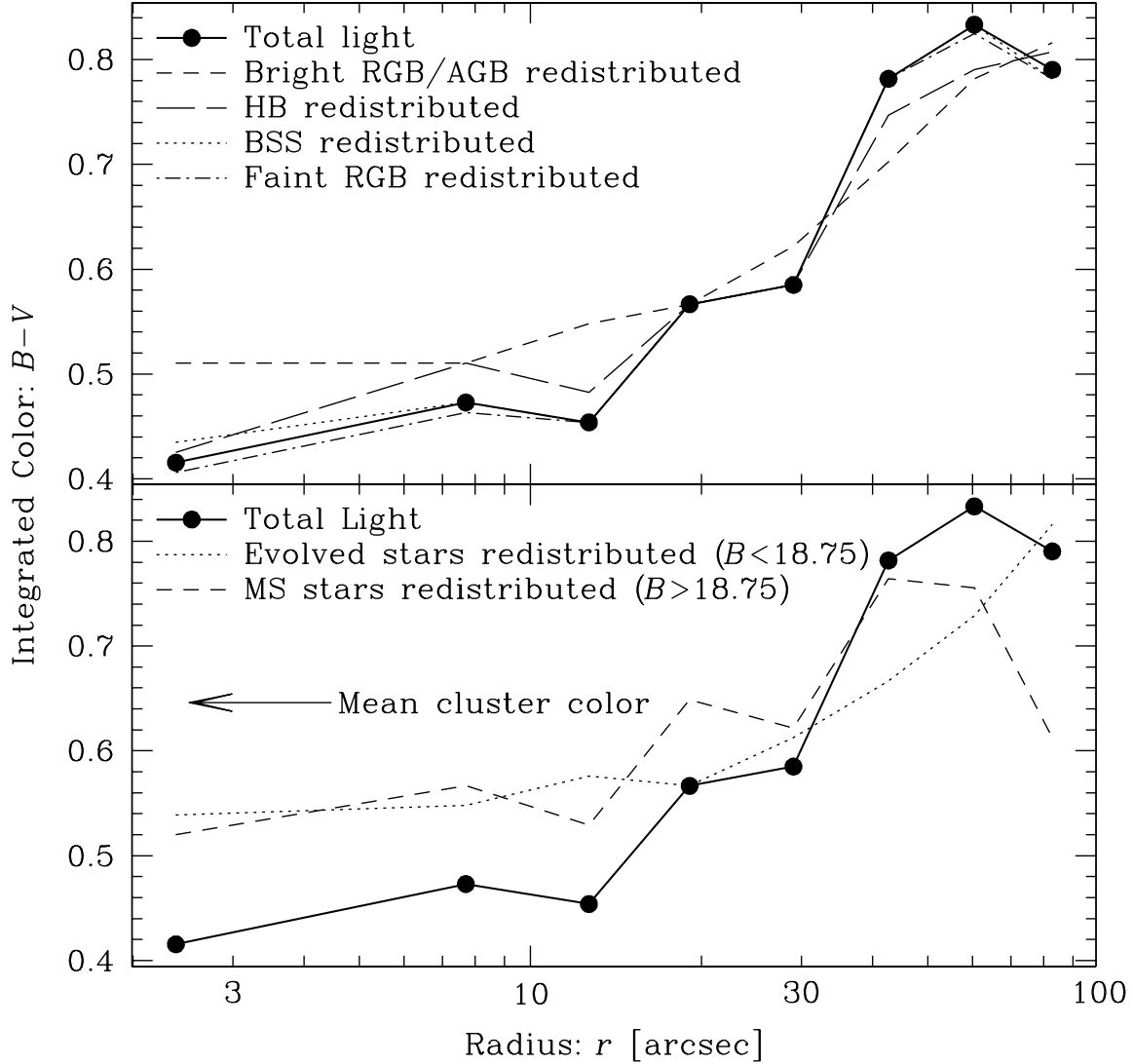}
\caption{The integrated $B-V$ color of M30's
starlight as a function of radius (bold solid line with dots).  Each of the
other curves illustrates the effect of ``uniformly'' redistributing the light
of a particular stellar type (see Sec.~{\it\ref{colorgrad}}).  The mean $B-V$
color of M30 over the area of the WFPC2 image ($r\lesssim2'$) is indicated in
the lower panel.  The integrated cluster starlight is about 0.4~mag bluer in
$B-V$ in the inner part than in the outer part.  Bright~RGB/AGB stars are
responsible for about 0.15~mag of this color gradient, and HB stars for about
0.05~mag.  The evolved stars taken together are responsible for about half of
the overall color gradient, while main sequence stars ($B>18.75$) are
responsible for the remainder of the gradient.  \label{cgrad.ps}}
\end{figure}

Figure~\ref{cgrad.ps} shows the integrated $B-V$ color of M30 as a function
of radius (bold line with dots in both panels).  Our measurements are in good
agreement with previous measurements of the color gradient of the overall
cluster light (\cite{piotto88}; \cite{burgbuat}) and may be
summarized as: ${\partial(B-V)}/{\partial\log(r)}=0.3$~mag/dex
with a color of $(B-V)_{\rm ref}\sim0.59$ at a radius of $r_{\rm ref}=30''$.

We next investigate which of M30's stellar types is/are responsible for the
observed color gradient.  If the color gradient were caused by a single
population of stars, the gradient would disappear when light from that
population in each band is redistributed ``uniformly'' throughout the
cluster.  We use the radial distribution of the integrated $B+V$ stellar flux
to define what is meant by ``uniform'' redistribution.  Before summing the
fluxes in the two~bands, the relative flux normalization is chosen such that
faint~RGB stars make an equal contribution in $B$ and $V$; these stars have
the same color as the mean color of the integrated cluster starlight over the
area of the WFPC2 image: $\langle{B-V}\rangle_{\rm M30}\sim0.65$.  For each
population (BSS, HB, bright~RGB/AGB, faint~RGB) in turn, we subtract the
actual flux of that population in every radial bin, and then redistribute the
total amount of flux removed in a uniform and identical fashion in both
bands.  The step of adding the light back in is critical as it dilutes any
remaining color gradient by the appropriate amount, thereby allowing us to
quantify the contribution of a given stellar type.

The effect of redistributing the light of a single evolved stellar population
on the overall color gradient is shown by the various curves in the upper
panel of Fig.~\ref{cgrad.ps}.  It is clear that bright~RGB/AGB stars affect
the color gradient more than any other evolved stellar type, but account for
only about a third of the observed gradient, or only
${\partial(B-V)}/{\partial\log(r)}=0.1$~mag/dex.  The observed lack of
smoothness in M30's integrated $B-V$ profile is a reflection of the
stochastic nature of the bright~RGB/AGB flux component: a relatively small
number of stars contributes a significant fraction of the total light.
Redistribution of the light of bright~RGB/AGB stars results in a considerably
smoother color gradient (short dashed line in upper panel of
Fig.~\ref{cgrad.ps}).  The population of blue HB stars, long a popular
suspect for the central bluing trend, do not affect the the overall color
gradient of the cluster light very much:
${\partial(B-V)}/{\partial\log(r)}=0.05$~mag/dex.  Another popular suspect,
the BSSs only affect the color by 0.02~mag in the innermost bin.

Although quite numerous near the center, the BSS sample is not a significant
contributor to the overall gradient.  A simple calculation verifies this:
bright RGB stars are about 3~times as numerous as BSSs in the center of M30
(see $F_{\rm BSS}$ in Sec.~{\it\ref{fbss}}); they have mean colors of
$\langle{B-V}\rangle_{\rm bright~RGB}\sim0.6$ and $\langle{B-V}\rangle_{\rm
BSS}\approx0$, a difference of $\Delta(B-V)=0.6$~mag; the typical BSS is
about 2~mag fainter than a bright RGB star (the ratio of their fluxes is
$\sim15$\%); bright RGB stars contribute about 50\% of the total cluster
light (see Sec.~\ref{sbprofile}); the fractional BSS portion of the
integrated stellar flux in the central bin is
$0.33\times0.15\times0.5=2.5$\%; thus, the concentration of BSSs in the
innermost bin is expected to affect the overall $B-V$ color by only
$0.025\times0.6=0.015$~mag. 

The lower panel of Fig.~\ref{cgrad.ps} shows that a smooth color gradient of
${\partial(B-V)}/{\partial\log(r)}=0.15$~mag/dex remains when the flux from
{\it all} evolved stars is redistributed.  This remaining color gradient must
be due to the unevolved stars (with $B>18.75$) in M30.  We estimate the
combined flux of all cluster MS stars---both resolved stars with
$V\lesssim21$ as well as those that are fainter than the detection
threshold---by subtracting the contribution of all evolved stars from the
total cluster flux in each bin.  By redistributing the MS light uniformly, we
verify that this population does indeed account for approximately half of the
overall color gradient observed (dashed line in the lower panel of
Fig.~\ref{cgrad.ps}).  The contribution of MS stars to M30's radial color
gradient is best explained in terms of mass segregation (\cite{pryor86};
\cite{bolte89}).  There is direct evidence that the fractional flux of MS
stars increases outwards across the eight~radial bins (see
Sec.~\ref{sbprofile} below).  Moreover, the mean color of the MS light is
expected to become redder with increasing radius because of an increasing
fraction of lower MS stars whose colors are redder than the cluster's mean
$B-V$ color of~0.65.  These faint red stars are mostly unresolved in the
WFPC2 images; the resolved MS stars tend to be the brighter ones just below
the MSTO and these have bluer colors (Fig.~\ref{bvbcmd}).

Djorgovski~\& Piotto (1993) suggested that M30's color gradient is caused by
depletion of giants in center of the cluster, while Burgarella~\& Buat (1996)
concluded that the color gradient cannot be explained by the evolved and
resolved stellar populations alone.  We find that bright~RGB/AGB stars are
responsible for only about a third of the overall color gradient and that
about half of the gradient is caused by stars below the MSTO.  There is a 5\%
chance that the central depletion of bright~RGB stars, and the associated
color gradient of ${\partial(B-V)}/{\partial\log(r)}=0.1$~mag/dex, is a
statistical fluke (Sec.~{\it\ref{center}}).  The contribution of HB stars to
the gradient is more likely to be a result of Poisson fluctuations.  The
color gradient due to the mass segregation of MS stars, about half the
overall gradient, is highly significant in a statistical sense.

\subsection{The Contribution of Various Stellar Types to the Integrated
Starlight} \label{sbprofile}

\begin{figure}
\plotone{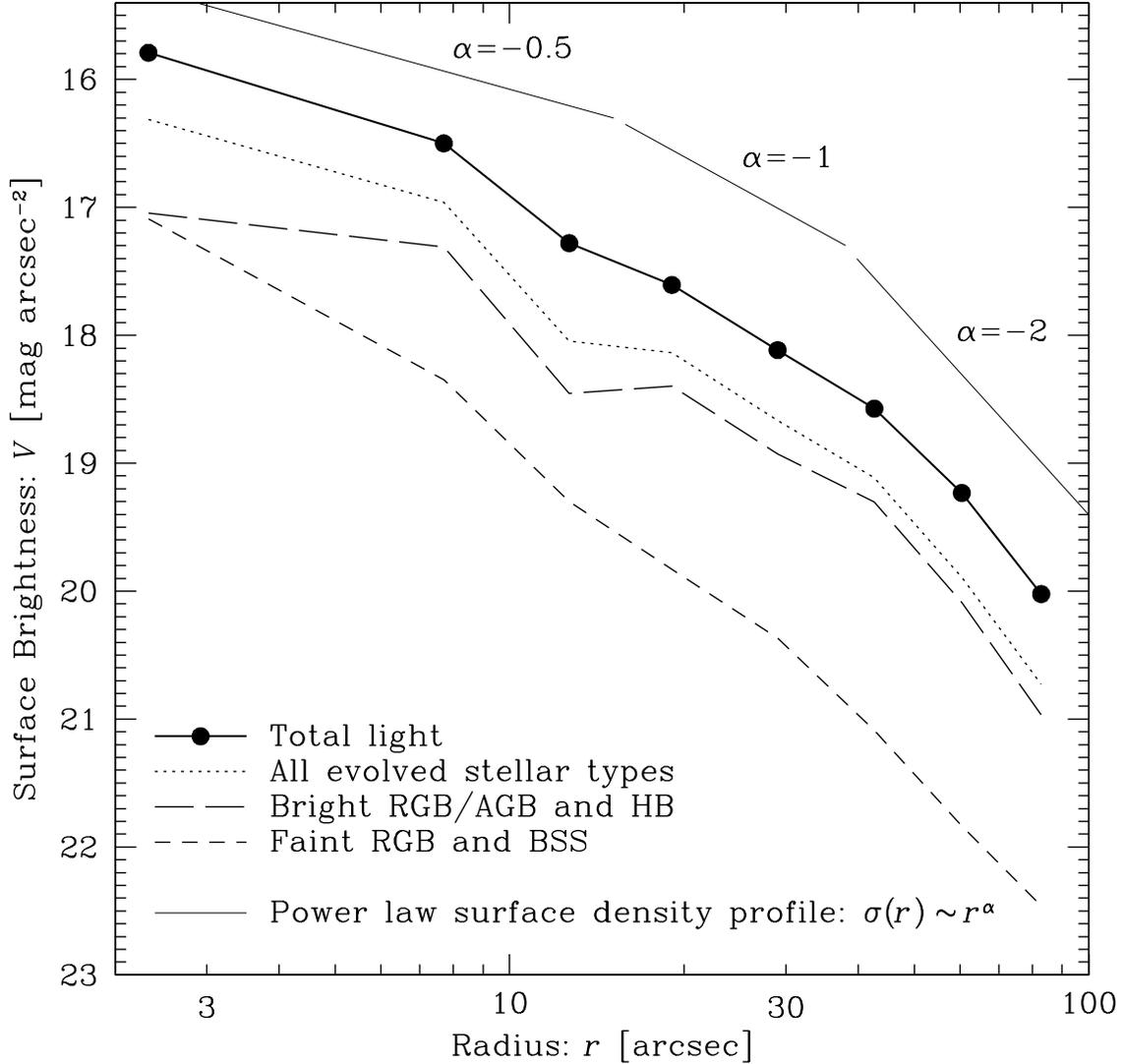}
\caption{The $V$ band surface brightness as a
function of radius in M30 for the integrated cluster starlight (bold solid
line with dots), as well the separate contributions of various stellar types:
all post main sequence stars, bright~RGB/AGB/HB stars, and faint~RGB/BSSs.
The main sequence component (total minus evolved stars) increases radially
outwards relative to the total light, and especially in relation to the
faint~RGB component, as a result of mass segregation.  Bright~RGB/AGB/HB
stars contribute 3--4~times as much flux as faint~RGB stars, except in the
innermost bin which is deficient in bright~RGB stars.  The profile of the
integrated light has a slope consistent with an $\alpha=-0.5$ power law for
$r<10''$ ($\alpha=-1$ for the faint~RGB light) and steepens to $\alpha=-2$
for $r>40''$.  \label{vmagarea}}
\end{figure}

Figure~\ref{vmagarea} shows the $V$ surface brightness as a function of
radius for various stellar populations as well as for the total cluster
light (bold line with dots).  Bright~RGB/AGB (defined by $B<16.6$ for this
study) and HB stars taken together account for most of the light of evolved
stars, especially beyond $r>20''$.  The central depletion of bright~RGB/AGB
stars relative to faint~RGB stars is apparent: the integrated
bright~RGB/AGB/HB flux is 4~times that of the faint~RGB flux in the outer
parts, whereas their fluxes are comparable in the innermost bin.  The MS
component (total flux minus integrated flux of all evolved stars) is about a
third of the total flux in the inner two~bins and increases to about half the
total flux in the outermost bin, likely a result of mass segregation.  The
``true'' degree of segregation of the MS component is slightly larger than
this observed fractional increase, since the effect is diluted by the central
depletion of bright~RGB/AGB stars.  For example, the ratio of integrated MS
flux to faint~RGB flux increases from~1.5--2 in the inner part
($r\lesssim10''$) to about~4--5 beyond $r>50''$.

The surface brightness profile of M30's integrated starlight displays
significant curvature in a log-log plot (Fig.~\ref{vmagarea}).  The slope
interior to $r=10''$ is about $\alpha=-0.5$, where $\alpha$ is the power law
index of the projected density profile: $\sigma(r)\propto{r}^\alpha$.  The
faint~RGB component's brightness profile slope in this same inner region is
$\alpha\sim-1$; this measure is unaffected by the central depletion of
bright~RGB stars and is in good agreement with the number-weighted estimate
of M30's density profile (\cite{letter}).  The profile slope for the total
light as well as for each stellar type steepens to about $\alpha=-2$ beyond
$r>40''$, the asymptotic power law slope of a King profile for $r>\!\!>r_{\rm
core}$ (\cite{king62}).

\subsection{Luminosity Function}\label{lf_vs_model}

The observed distribution of red giant luminosities in a globular cluster,
while relatively easy to measure, can serve as a direct and powerful
diagnostic of the rate of fuel consumption as a function of the evolutionary
stage of the star from the beginning of the post-main-sequence phase
(development of hydrogen burning shell) to the tip of the RGB phase (helium
flash).  The luminosity of an RGB star is expected to be determined
principally by the mass of its helium core (\cite{refsdal}).  Detailed
stellar evolutionary models do indeed confirm that the gross shape of the LF
of red giants (those significantly brighter than the subgiant branch) is
largely independent of the metallicity and age of the cluster
(\cite{bergbvan}).

\begin{figure}
\plotone{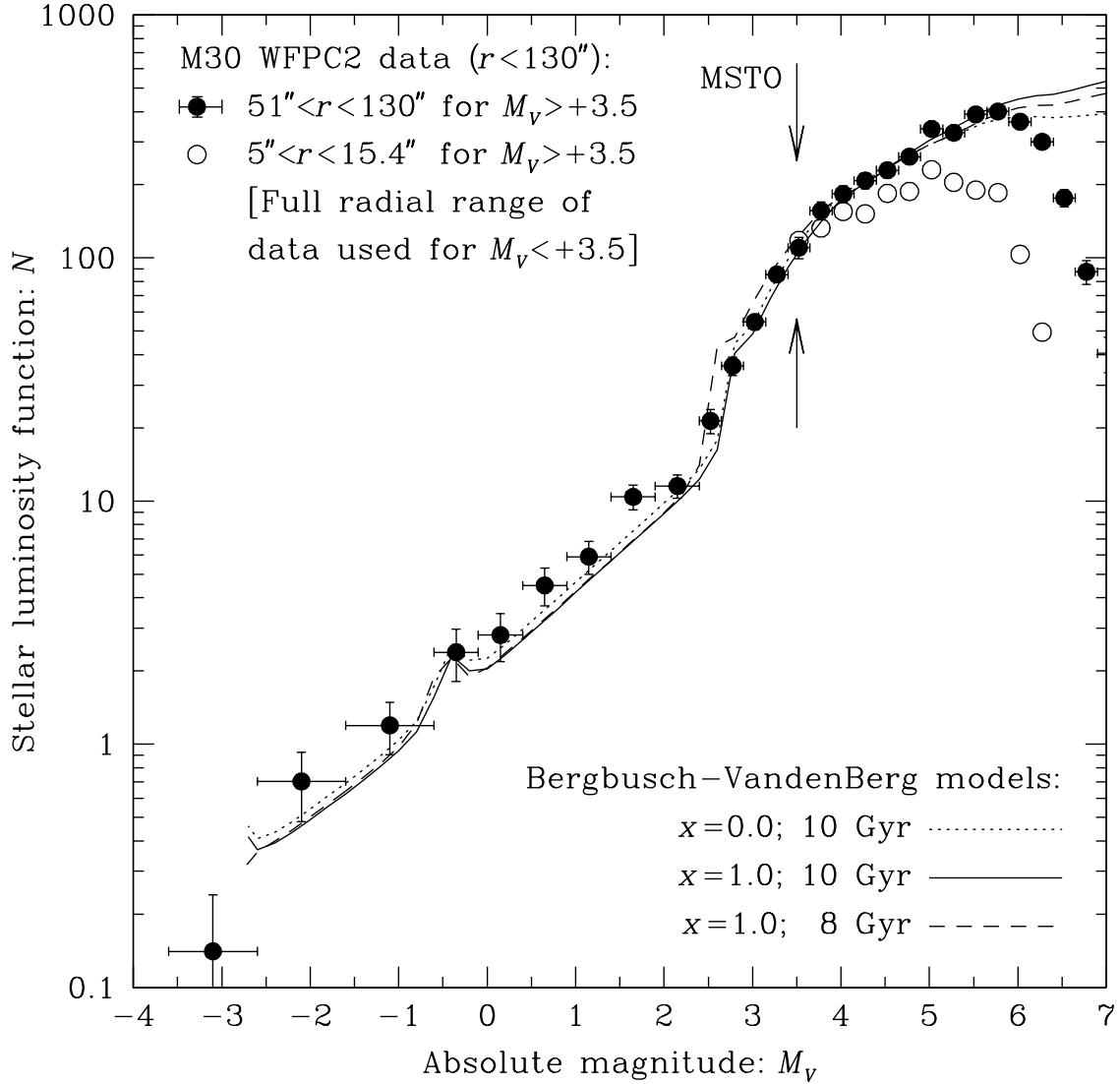}
\caption{The $V$-band stellar luminosity function
of M30 (bold dots with Poisson error bars and bin width markers) compared to
the Bergbusch \& VandenBerg (1992) models for 8~Gyr and 10~Gyr metal-poor
populations for various mass function slopes $x$ (lines).  The models are
normalized to the observed number of stars in the range $M_V=+4.5$ to +5.0.
The data points for bright ($M_V<+2.5$) stars are based on the entire WFPC2
mosaic; the faint star data (bold dots) are based only on the relatively
sparse outer radial bins ($r\gtrsim50''$).  The turnover in the counts for
$M_V>6$ is due to incompleteness; the effect of incompleteness is stronger
and sets in at brighter magnitudes closer to the cluster center (open
circles).  The models underpredict the number of bright red giants by 30\%.
\label{theolum}}
\end{figure}

Figure~\ref{theolum} compares the observed M30 $V$-band stellar LF to model
isochrones computed by Bergbusch~\& VandenBerg (1992) with abundances of
$\rm[Fe/H]=-2.03$, [O/Fe]$=+0.70$, and $Y=0.235$.  The adopted composition is
consistent with abundance estimates for the cluster (cf.~\cite{zinnwest};
\cite{carretta}; \cite{Sandquist}).  An apparent distance modulus of
$(V-M_V)_{\rm app}=15.10$, based on fitting subdwarfs with {\it Hipparcos\/}
distance measurements to the cluster MS (\cite{reid}) and line-of-sight
reddening of $E_{B-V}=0.05$ (\cite{zinn80}; \cite{burs}; \cite{richer}), is
adopted.  Most AGB stars (all but the reddest ones) and all HB stars are
excluded in computing M30's LF since such stars are absent from the
Bergbusch~\& VandenBerg model; the remaining AGB stars make a negligible
perturbation on the LF, as do BSSs ($\lesssim10$\% effect for the bins around
$M_V=+2$ to +3).  No completeness correction has been applied; incompleteness
is unimportant across most of the brightness range shown in
Fig.~\ref{theolum} ($M_V<+5.5$), as the stars fainter than the MSTO (solid
dots) are drawn only from the outer two~annuli ($51\arcsec<r<130\arcsec$)
where crowding is not too severe.  The relatively rare bright stars
comprising the $M_V<+2.5$ portion of the LF are counted over the full WFPC2
image ($r<130''$); the bright and faint parts of the LF are matched to each
other using counts of all stars with $V<V_{\rm MSTO}$ ($M_V<+3.5$).

The Bergbusch~\& VandenBerg model LFs shown in Fig.~\ref{theolum} are
normalized to match the data over the range $19.5\lesssim V\lesssim20.0$
($+4.5\lesssim{M_V}\lesssim+5.0$).  The 10~Gyr and 8~Gyr isochrones provide
acceptable fits to M30's MS stars in the range $+3.5<M_V<+5.5$
(incompleteness sets in for $M_V\gtrsim+6$ or $V\gtrsim21$) for initial mass
function slopes in the range $x=0$ to $x=1$, where $x$ is defined by the
usual relation: $dN(M)\propto M^{-(1+x)}dM$.  The portion of the LF with
$M_V<+5.5$ corresponds to a very narrow range in stellar mass
($0.73\,M_\odot<M<0.90\,M_\odot$ for a 10~Gyr isochrone) so that the shape of
this portion of the model LF is almost independent of the mass function slope
$x$.  Based on studies extending further down the MS beyond $r>2'$
(\cite{bolte94}; \cite{Sandquist}), a mass function slope of $x=1$ is quite
reasonable for the $r\sim1'\>$--$\>2'$ region of the cluster from which the
faint stars in Fig.~\ref{theolum} are drawn.  While the isochrones are
normalized to the MS portion of M30's LF, the 10~Gyr isochrone also fits the
subgiant portion well.  The MSTO luminosity for the 8~Gyr isochrone is
noticeably brighter than in M30, and this is manifested as an excess in the
predicted number of subgiants across the range $+2.5<M_V<+3.5$.

The number of RGB stars with $M_V<+2$ ($V<17.1$) found in M30 is~220, 30\%
higher than predicted by the appropriately normalized model LFs (170).  The
deficit is significant at the $3.8\sigma$ level.  The effect is independent
of RGB luminosity for $M_V<+2$.  Earlier ground-based studies have noted a
similar mismatch between observation and standard stellar evolutionary models
in the $r>2'$ region of M30 (\cite{bolte89}; \cite{Sandquist}).  The ``RGB
excess'' phenomenon is not unique to M30---the metal poor clusters M5
(\cite{sand96}) and M13 (Paper~VI) display an RGB excess relative to
theoretical models, but unlike the case of M30, the excess in these
two~clusters is restricted only to giants brighter than the RGB bump or HB
level ($M_V\lesssim0$).

One possible way of resolving the discrepancy between M30's observed RGB LF
and theoretical LFs is by invoking the effect of internal core rotation in
stellar models (\cite{larson}; \cite{vdb98}).  This tends to decrease the
rate of RGB evolution (by expanding the core and lowering the shell
temperature) and to thus increase the number of giants in any luminosity bin
at a given instant.  Core rotation models have only been constructed,
however, for giants fainter than $M_V=0$; thus, these models do not directly
address the RGB excesses observed in M5 and M13.  A more comprehensive
solution based on ``deep mixing'' has recently been explored by
\cite{langer}.  Deep mixing has already been invoked to explain: (1)~chemical
abundance anomalies in some bright cluster giants in terms of processing of
envelope material in the hotter energy generating shell (cf.~\cite{kraft});
and (2)~anomalies in the HB color (second parameter problem) and brightness
in terms of introduction of helium rich material into the envelope
(\cite{sweigart}).  Langer et~al.\ propose that the ``extra'' fuel mixed into
the hydrogen burning shell might also account for the overabundance of RGB
stars seen in certain clusters.

The observed excess of red giants relative to the standard model is not an
artifact resulting from incompleteness in the faint star counts.  As
discussed above, the shape of the MS LF for $+3.5<M_V<+5.5$ is a good match
to model LFs, so any incompleteness in the star counts would have to be
independent of apparent magnitude over this range ($18.5<V<20.5$).  Our image
simulations indicate that the sample is complete down to $V\sim21$ over the
$r\sim1'\>$--$\>2'$ region from which this part of the LF is constructed
(Sec.~\ref{acccomp}).  Moreover, incompleteness tends to be a strong function
of apparent magnitude; this is clearly seen in the crowded inner region
($5\arcsec<r<15\arcsec$) of M30 where incompleteness sets in at brighter
magnitudes (open circles in Fig.~\ref{theolum}; see also Fig.~\ref{vlum}).

The conclusion about M30's RGB excess is essentially independent of the
adopted cluster distance modulus $(m-M)$ and choice of isochrone age $t$
(Fig.~\ref{theolum}).  For {\it any\/} reasonable value of $(m-M)$, $t$ can
be chosen so that the model LF (\cite{bergbvan}) fits the MS and subgiant
``bump'' portions of M30's LF, but the model underpredicts the number of RGB
stars.  Even if the assigned value of $t$ is too large (too small) for a
given distance modulus, the result is an apparent excess (deficiency) in
M30's subgiant population relative to the model, provided the model LF is
normalized to the MS portion of M30's LF; the degree of M30's RGB excess
though is unaffected by such a mismatch between $t$ and $(m-M)$.  In spite of
recent improvements in the measurement of subdwarf parallaxes with {\it
Hipparcos\/}, the distance modulus of M30 remains uncertain at the level of
$\Delta(m-M)\gtrsim0.2$~mag, corresponding to an uncertainty in the cluster
age of $\Delta{t}\gtrsim2$~Gyr (\cite{reid}; \cite{gratton}; \cite{pont};
\cite{Sandquist}).  The distance scale we have adopted, $(V-M_V)_{\rm
app}=15.10$ (\cite{reid}) happens to be at the high end of this range, so
that the corresponding best-fit age, $t\sim10$~Gyr, is lower than in other
studies.

\section{Conclusions}

\begin{itemize}

\item[{\bf 1.}]{\it Hubble Space Telescope\/} Wide Field/Planetary Camera~2
images of the dense globular cluster M30 (NGC~7099) in the F555W, F439W, and
F336W filters have been analyzed.  Accurate stellar positions and photometry
in the (Johnson) $UBV$ bands are presented for 9507~stars (and $BV$
photometry for an additional 433~stars) within a projected distance of
130\arcsec\ from the cluster center.  Color-magnitude diagrams based on the
$UBV$ bands are presented showing clearly distinguished sequences of stellar
types: red giant branch, (blue) horizontal branch, subgiant branch, blue
straggler sequence, main sequence turnoff, and main sequence.  Typical
photometric errors for main sequence stars ($B=19\>$--$\>$20) range from
$\sim0.1$~mag for the central 10\arcsec\ of M30 to $\sim0.05$~mag in the rest
of the cluster.  Incompleteness sets in a little fainter than the main
sequence turnoff ($V\gtrsim19$) for $r<5\arcsec$ but the sample is complete
to progressively fainter magnitudes at larger radii ($V\sim20.5$ for
$r\gtrsim20''$).

\item[{\bf 2.}]An unusual star is found $1\farcs2$~SSW of the cluster center.
Its very blue color ($B-V=-0.97$, $B=18.6$) indicates that it might possibly
be a cataclysmic variable.

\item[{\bf 3.}]Forty-eight blue straggler candidates are identified in the
cluster on the basis of a ($B-V$,~$B$) color-magnitude diagram; the BSS
classification for some of these is tentative.  The specific frequency of BSS
in M30, $F_{\rm BSS}\equiv N({\rm BSS})/N(V<V_{\rm HB}+2$, is~$0.39\pm0.08$
in the inner 10\arcsec\ and $0.14\pm0.02$ over the full WFPC2 image
($r<130\arcsec$).  The central BSS specific frequency in M30 is the highest
measured for any globular cluster to date.  The BSSs are strongly centrally
concentrated relative to other evolved stars ($>99\%$ significance).  The
observed M30 BSS luminosity function is compared to theoretical predictions:
although the number of BSS is too small to draw definite conclusions, the
collisional formation model seems to be a better match to the data than the
primordial binary merger model. 

\item[{\bf 4.}]The abundance of bright red giant branch stars and asymptotic
giant branch stars with $V\lesssim16$ (relative to fainter RGB stars and
subgiants) appear to be a factor of~2--3 lower in the central 15\arcsec\ of
M30 compared to further out in the cluster (2--2.5$\sigma$ effect).
Horizontal branch stars, the evolutionary descendants of bright RGB stars,
show no significant central depletion relative to the distribution of faint
RGB stars and subgiants.

\item[{\bf 5.}]The $B-V$ color of the integrated cluster light in M30 varies
from 0.82 around $r\sim1'$ to 0.45 within $r<10''$, a radial color gradient
of ${\partial(B-V)}/{\partial\log(r)}=0.3$~mag/dex.  The central deficiency
of bright red giant/asymptotic giant branch stars is responsible for about a
third of the color gradient; all evolved stellar populations taken together
(RGB, AGB, HB, BSS, subgiants) are responsible for about half of the overall
color gradient.  Mass segregation of faint red main sequence stars results
in a smooth $B-V$ color gradient of about 0.15~mag/dex, corresponding to half
the observed gradient.

\item[{\bf 6.}]The $V$-band surface brightness profile of the integrated
starlight in M30 has a power law slope of $\alpha=-0.5$ in the inner
10\arcsec, and the slope is $\alpha=-1$ for the light of faint~RGB stars
alone; the latter is identical to the number-weighted measure of the
projected density profile slope found in \cite{letter}.  The difference
between the central slopes of the overall vs faint~RGB light is a result of
the central depletion of bright~RGB stars.  The integrated surface brightness
profile slope steepens to $\alpha=-2$ for $r>40\arcsec$.  Main sequence stars
contribute nearly 40\%, bright~RGB/AGB/HB stars with $V\lesssim16$ contribute
30\%, and faint~RGB stars contribute about 30\% of the overall cluster
brightness in the central $5''$ of M30; the corresponding fractions are 50\%,
40\%, and 10\% at larger radii ($r\sim1'$).

\item[{\bf 7.}]The stellar luminosity function in the inner $2'$ of M30 shows
an anomalously high abundance of red giants (relative to subgiant, main
sequence turnoff, and main sequence stars), as has been previously noted
further out in the cluster.  Giants with $M_V<+2$ in M30 are overabundant by
30\% (3.8$\sigma$ significance) relative to the prediction of suitably
normalized standard stellar evolution models.

\end{itemize}

\bigskip
\bigskip
\acknowledgments
We would like to thank Mike Bolte, Karl Gebhardt, Josh Grindlay, Marc
Pinsonneault, and Eric Sandquist for useful suggestions.  This project was
supported in part by an Alfred P.\ Sloan Foundation fellowship (PG), in part
by a National Science Foundation Graduate Fellowship (ZTW), and in part by
NASA through Grant No.\ NAG5-1618 from the Space Telescope Science Institute,
which is operated by the Association of Universities for Research in
Astronomy, Inc., under NASA Contract No.\ NAS5-26555.

\clearpage
\appendix

\centerline{\bf Appendix: Table of Stellar Photometry and Positions}

\medskip
The complete version of Table~\ref{datatable} contains (Johnson) $U$-,
$B$-, $V$-band photometry for 9507~stars with $V\lesssim21$ in the central
5.1~arcmin$^2$ of M30 ($r<130''$) covered by the {\it HST\/} WFPC2 image, and
$B$ and $V$ for an additional 433~stars that were {\it not\/} matched in the
$U$ band; a subset containing the 40~brightest stars within $r<10''$ of M30's
center is presented in the paper.  A computer-readable version of the
complete table is available in the electronic edition of the Journal by link
to a permanent database.  It may also be obtained via {\tt anonymous ftp} as
follows:

\begin{itemize}

\item[$\bullet$]{\tt ftp eku.sns.ias.edu} (login as: {\tt anonymous})

\item[$\bullet$]{\tt cd pub/GLOBULAR\_CLUSTERS/m30-ppr7}

\item[$\bullet$]{\tt get m30fulltab1.txt}

\end{itemize}

\noindent
or by contacting P.G..

Details of the photometric procedure, a description of the astrometric
convention, and estimates of the completeness and photometric accuracy may be
found in Sec.~\ref{dataproc}, \ref{astrom}, and \ref{acccomp}, respectively.
We recommend that the astrometry in Table~\ref{datatable} be matched
empirically against other data sets to correct for possible systematic errors
in the scale, translation, and rotation of the coordinate system we have
adopted.

\end{document}